\documentclass[12pt,preprint]{aastex}

\def\Ho {H_{0}}
\def\eg   {e.g.~}

\begin{document}

\title{THE MEGAMASER COSMOLOGY PROJECT.\uppercase\expandafter{\romannumeral8}. \\ 
A GEOMETRIC DISTANCE TO NGC 5765b 
 }

\author{F. Gao\altaffilmark{1,2,3}, J. A. Braatz\altaffilmark{2},  M. J. Reid\altaffilmark{4}, K. Y. Lo\altaffilmark{2},  J. J. Condon\altaffilmark{2}, 
            C. Henkel\altaffilmark{5,6}, C. Y. Kuo\altaffilmark{7}, C. M. V. Impellizzeri\altaffilmark{2,8}, D. W. Pesce\altaffilmark{9} AND W. Zhao\altaffilmark{1}} 
            
\altaffiltext{1}{Key Laboratory for Research in Galaxies and Cosmology, Shanghai Astronomical Observatory, Chinese Academy of Science, Shanghai 200030, China}
\altaffiltext{2}{National Radio Astronomy Observatory, 520 Edgemont Road, Charlottesville, VA 22903, USA}
\altaffiltext{3}{Graduate School of the Chinese Academy of Sciences, Beijing 100039, China}
\altaffiltext{4}{Harvard-Smithsonian Center for Astrophysics, 60 Garden Street, Cambridge, MA 02138, USA}
\altaffiltext{5}{Max-Planck-Institut f\"ur Radioastronomie, Auf dem H\"ugel 69, 53121 Bonn, Germany}
\altaffiltext{6}{King Abdulaziz University, P.O. Box 80203, Jeddah, Saudi Arabia}
\altaffiltext{7}{Academia Sinica Institute of Astronomy and Astrophysics, P.O. Box 23-141, Taipei 10617, Taiwan}
\altaffiltext{8}{Joint Alma Office, Alsonso de Cordova 3107, Vitacura, Santiago, Chile }
\altaffiltext{9}{Department of Astronomy, University of Virginia, P.O. Box 400325, Charlottesville, VA 22904, USA }

\begin{abstract}
As part of the Megamaser Cosmology Project (MCP), here we present a
new geometric distance measurement to the megamaser galaxy NGC
5765b. Through a series of VLBI observations, we have confirmed the
water masers trace a thin, sub-parsec Keplerian disk around the
nucleus, implying an enclosed mass of 4.55 $\pm$ 0.40
$\times~10^{7}M_\odot˜$. Meanwhile, from single dish monitoring of
the maser spectra over two years, we measured the secular drifts of
maser features near the systemic velocity of the galaxy with rates
between 0.5 and 1.2 km~s$^{-1}$~yr$^{-1}$. Fitting a warped, thin disk
model to these measurements, we determine a Hubble Constant $\Ho$ of
66.0 $\pm$ 6.0 km~s$^{-1}$ Mpc$^{-1}$ with the angular-diameter
distance to NGC 5765b of 126.3 $\pm$ 11.6 Mpc.

Apart from the distance measurement, we also investigate some physical
properties related to the maser disk in NGC 5765b. The high-velocity
features are spatially distributed into several clumps, which may
indicate the existence of a spiral density wave associated with the
accretion disk. For the red-shifted features, the envelope defined by
the peak maser intensities increases with radius. The profile
of the systemic masers in NGC 5765b is smooth and shows almost no
structural changes over the two years of monitoring time, which
differs from the more variable case of NGC 4258.

\end{abstract}

\section{INTRODUCTION}

Recent measurements of the cosmic microwave background (CMB)
anisotropies have provided dramatic confirmation of the ``standard''
$\Lambda$CDM model (e.g. Hinshaw et al. 2013). However, degeneracies
between parameters such as the dark energy equation of state, the
Hubble constant and the neutrino mass limit the determination of each
quantity from the CMB data alone. One way to break this degeneracy is
an independent and high precision measurement of $\Ho$ (better than a
few percent), as pointed out in Hu (2005). This motivation has
reinvigorated efforts to measure $\Ho$ and has driven the development
of better methods of measuring $\Ho$ over the past few
years. For a general review on this topic, please see Freedman \&
Madore (2010) and references therein.

So far, measurements of $\Ho$ have achieved about 3 \%
precision. Riess et al. (2011) derived $\Ho$ = 73.8 $\pm$
2.4~km~s$^{-1}$ Mpc$^{-1}$ by using both Type Ia supernovae and
Cepheids. Meanwhile, by improving the calibration of the
period-luminosity (P-L) relation of Cepheids using mid-infrared
observations, Freedman et al. (2012) obtained $\Ho$ = 74.3 $\pm$ 2.1
km~s$^{-1}$ Mpc$^{-1}$ from the Carnegie Hubble Program. 

On the other hand, with the flat geometry assumption, the
model-dependent prediction of $\Ho$ from the most recent Planck result
has a lower $\Ho$ of 67.8 $\pm$ 0.9~km~s$^{-1}$ Mpc$^{-1}$ (Planck
Collaboration XIII 2015), while the recent complete WMAP nine year
observations constrain $\Ho$ to be 70.0 $\pm$ 2.2~km~s$^{-1}$
Mpc$^{-1}$ (Hinshaw et al. 2013). Both of these predictions are in
tension with the directly measured $\Ho$ from Type Ia supernovae and
Cepheids, as mentioned above. It is not clear so far whether this
tension means new physics are needed, adding to the basic six parameter
$\Lambda$CDM model, or any unknown systematic errors exist in either
of the measurements. In either case, a direct measurement of $\Ho$
independent of the methods mentioned above is highly desired.

As one of the direct distance measurement methods, the ``disk maser''
method has demonstrated its capability for precise distance
measurement through the work on the archetypal disk megamaser galaxy
NGC 4258 (\eg Herrnstein et al. 1999; Humphreys et al. 2013). This
method utilizes the 22 GHz water maser emission, which traces a
sub-parsec thin accretion disk in Keplerian rotation at the center of
a galactic nucleus, to obtain the angular-diameter distance. The maser distance 
to NGC 4258 provides an anchor point for the distance ladder using Cepheids 
and type Ia supernovae in Riess et al. (2011).  When applied to more distant 
galaxies well into the Hubble flow, the disk maser method provides a simple, 
one-step measurement of $\Ho$ that doesn't involve secondary calibrations, 
which may introduce extra systematic errors.

With this motivation, the Megamaser Cosmology Project (MCP) aims to
make an independent measurement of $\Ho$ to a few percent uncertainty
by measuring the angular-diameter distances to disk megamaser galaxies
well into the Hubble flow (between 50 and 200 Mpc). So far, the MCP
group has published results on three galaxies, UGC 3789 with $\Ho$ =
68.9 $\pm$ 7.1~km~s$^{-1}$ Mpc$^{-1}$ (Reid et al. 2013), NGC 6264
with $\Ho$ = 68.0 $\pm$ 9.0~km~s$^{-1}$ Mpc$^{-1}$ (Kuo et al. 2013)
and NGC 6323 with $\Ho$ = 73$^{+26}_{-22}$~km~s$^{-1}$ Mpc$^{-1}$ (Kuo
et al. 2015). Measurement of distances for other megamaser-host
galaxies and the search for more suitable distance measurement
candidates are ongoing.

\begin{figure}[h]
\rotate
\epsscale{1.00}
\plotone{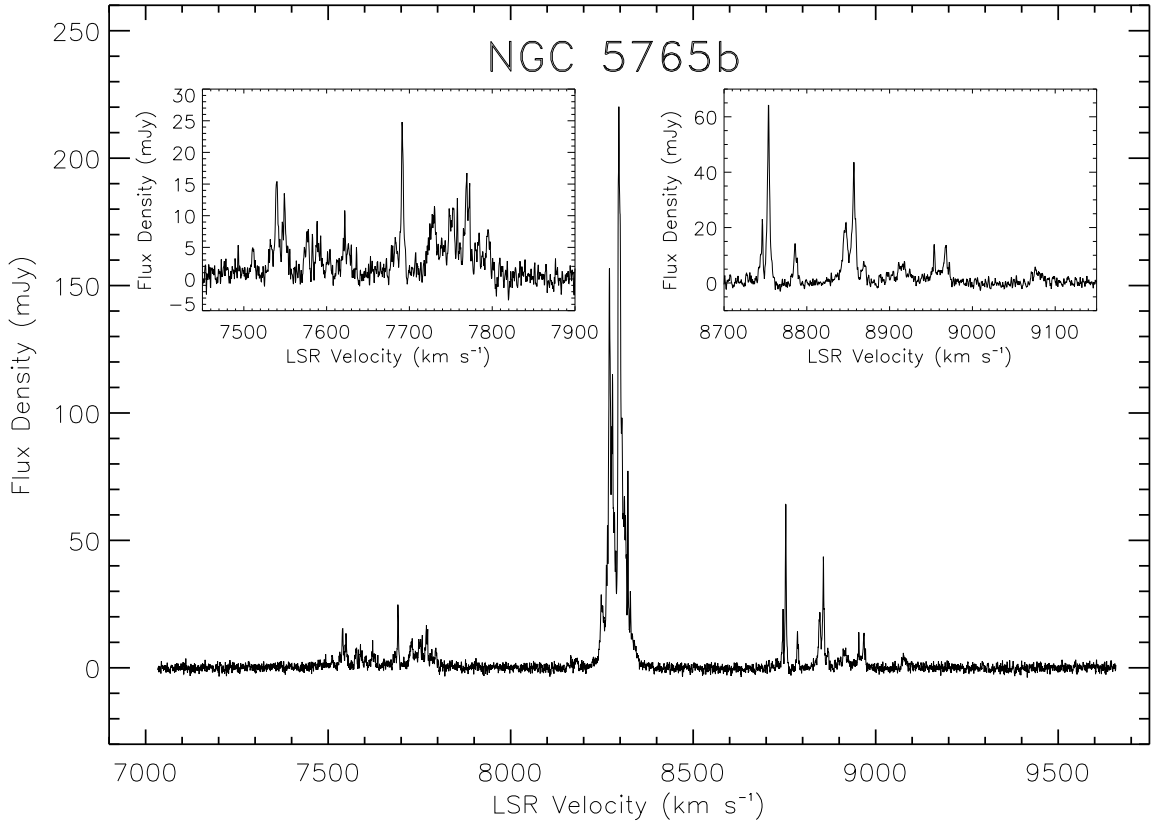} 
\caption{\footnotesize 
A representative spectrum of the 22 GHz water maser in NGC 5765b,
observed on 2014 April 06 by the GBT. The LSR recession velocity of
the galaxy is 8345 km~s$^{-1}$, as obtained from NED. The two insets
show zoomed-in details of the high-velocity masers.
\label{figure:gbtspec}
}
\end{figure}

As part of the MCP effort, here we present a new geometric distance
measurement to the megamaser galaxy NGC 5765b.  NGC 5765b is an Sa-b
galaxy in a galaxy pair together with NGC 5765a. It harbors a Seyfert
2 active galactic nucleus. The 22 GHz water maser emission was
discovered in 2012 February during our survey observations using the
Robert C. Byrd Green Bank Telescope (GBT)\footnote{The Green Bank
  Telescope is a facility of the National Radio Astronomy
  Observatory.}. Fig.~\ref{figure:gbtspec} shows a representative GBT
single-dish spectrum, which has a typical disk maser profile showing
three sets of emission lines, indicating a rotation speed about 600
km~s$^{-1}$. The recession velocity of NGC 5765b is $\sim$ 8345
~km~s$^{-1}$, which is well into the Hubble flow. In this paper, we
use optical velocity defined in the LSR frame throughout. To convert
to Heliocentric for NGC 5765b, subtract 12 km~s$^{-1}$ from the LSR
velocity.

This paper focuses solely on the distance measurement
to NGC 5765b and some physical properties of the maser disk specific to
this galaxy. In a forthcoming paper, we will summarize the results from 
the four published galaxies measured by the MCP, present a combined $\Ho$ 
measurement, and summarize plans to complete the project. 
We will compare $\Ho$ from masers to that measured from other methods,
and present updates on constraints of other cosmological parameters in the 
final summary paper.

This paper is arranged as follows: we report the VLBI observations and
results in Section 2. Section 3 describes our single dish monitoring
observations and the acceleration measurements. In Section 4 we
describe the distance measurement with a Bayesian approach and give
the fitting results and error budgets. In Section 5, first we discuss
some essential requirements for good distance measurement based on the
case of NGC 5765b. Then we focus on some physical properties seen in
NGC 5765b, including the clumpiness of the red-shifted features, the
correlation between maser flux and radius seen in the red-shifted
features, and the variability of the systemic maser profile compared
to the case of NGC 4258. We present our conclusions in Section 6.

\section{ VLBI OBSERVATIONS }

To map the spatial distribution of maser spots in NGC 5765b, we
conducted a series of VLBI observations between 2012 April and 2014
January, with the 10 antennas of the Very Long Baseline
Array(VLBA)\footnote{The VLBA is a facility of the National Radio
Astronomy Observatory, which is operated by the Associated
Universities, Inc. under a cooperative agreement with the National
Science Foundation (NSF).}, augmented by the 100m GBT, the
Effelsberg 100m telescope (EB), and the Karl G. Jansky Very Large Array
(VLA). We list all the basic observing information in
Table~\ref{table:VLBI observation}, including project code, date
observed, duration, antennas used, beam size, sensitivity, and the
observing mode.
    
\subsection{VLBI absolute position measurement for NGC 5765b}
       
Following the detection of maser emission in NGC 5765b by the GBT, we
conducted phase referencing VLBI observations (project code BB313AC
and BB321Y5) to measure the maser's sky position. Based on our
experience, to maintain good phase coherence and calibration, we require
an absolute position measurement for the NGC 5765b maser good to $\sim$10
mas (see details in Appendix~\ref{a:positionaccuracy}). We
followed the observation setup and data reduction process for
phase-referencing observations as described in Reid et al. (2009). Two
nearby quasars, J1448+0402 and J1458+0416, were chosen as phase
reference calibrators and we cycled the observations between these 2
calibrators and NGC 5765b.
	
The data reduction was done with NRAO's Astronomical Image Processing
System (AIPS) software. From BB313AC we derived the sky position of
the peak maser emission at velocity 8294.0 km~s$^{-1}$, which is RA=
$14^h50^m51^s.51950$ and DEC= $+05^{\circ}$06'52".2502. We use this
position in correlating later VLBI observations. The position
difference we measured between BB313AC and BB321Y5 is 0.1 mas in RA
and 0.7 mas in DEC, which is quoted as the measured absolute position
uncertainty. We list the sky positions for all these 3 sources in
Table~\ref{table:VLBI source position}.

       	\subsection{New VLBI capability starting 2013 February}
    
In general, our ability to measure the relative positions of megamaser spots with VLBI 
is limited by sensitivity, with the position uncertainty of each maser spot 
approximately given by
0.5$\Theta$/(S/N), in which $\Theta$ is the synthesized beam size, and
S/N is the signal-to-noise ratio. To reduce the noise level would
require a longer integration time and/or larger array collecting
area. The typical intensity of maser emission in our targets is a few
tens of mJy and line widths are $\sim$ 2 km~s$^{-1}$. Based on our
experience, mapping 10 mJy maser line requires a $\sim$ 10-hour VLBI
observation using the VLBA+GBT+EB. The case of NGC 6323 further
demonstrates the necessity of VLBI sensitivity, in which thirteen
12-hour VLBI tracks were conducted to map the $<$10 mJy systemic
masers in NGC 6323 (Kuo et al. 2011), but in the end their
sensitivity-limited VLBI measurement is still the limiting factor for
determining $\Ho$ as concluded in Kuo et al. (2015).
    
Our VLBI observations benefit from two significant upgrades which were
implemented in 2013 February:
    
1) The VLBI observations were joined by the VLA in the phased-array mode,
following the completion of the expanded VLA project. The phased-VLA
is a key component in our VLBI observations as it improves our
sensitivity by up to 40\%.
    
2) As part of the VLBA sensitivity upgrade project, the new
ROACH Digital Backend(RDBE), together with the Mark 5C recording
system, replaced the old VLBA legacy system for all stations included
in our observations. The new recorder increases the data recording
rate from 512 Mbit~s$^{-1}$ to 2 Gbit~s$^{-1}$, allowing us to use two
128 MHz bands to cover the whole maser spectrum contiguously, with
dual polarization and 2-bit sampling. For comparison, with the
recording rate of 512 Mbit~s$^{-1}$ and maximum bandwidth of each
baseband limited to 16 MHz as before, we could barely cover the whole
spectrum of NGC 5765b with the eight available basebands, as both the
blue-shifted and red-shifted features span over 20 MHz in
frequency. So the new system increases the sensitivity by increasing
observation efficiency. For data correlation, in addition to the
default ``continuum-like" correlation pass with 0.5 MHz channel size
over the whole band, we obtained a second correlation pass with three
narrow bands of 32 MHz bandwidth centered on the three sets of maser
lines using the new zoom-band mode offered in the DiFX software
correlator. The spectral resolution in our raw data is 25 kHz ($\sim$
0.3 km~s$^{-1}$), which is comparable to that of our GBT single dish
spectrum (24.4 kHz). This is 5 times finer than what we could achieve
with the legacy system and it allows higher accuracy of matching
between the single dish and VLBI spectra.
        
Except for the two VLBI tracks conducted in phase referencing mode
when measuring the absolute position of NGC 5765b, all other VLBI
observations were observed in the ``self-cal'' mode, in which we use
the strongest maser line as the phase calibrator. Our observation
setup and data reduction process are similar to previous MCP
observations, so please refer to earlier papers (Reid et al. 2009; Kuo
et al. 2013) for more details. Here we only describe differences from
earlier work.
    
For the phased VLA, we need to ``phase it up'' prior to each VLBI
scan. The ``phase up'' was done by observing J1448+040 ($>$ 200 mJy,
1.1$^{\circ}$ away from NGC 5765b) for 1 minute between each VLBI scan. 
To maintain good phase
coherence for the VLA, we set the maximum VLBI scan length to be 10
mins with the VLA in B or C configuration. We also added a flux
calibration scan for the VLA to get the combined system temperature
for the phased array. For all self-calibration tracks, we used the
hourly observed delay calibrator to solve for the change of multi-band
delay in time rather than using geodetic blocks as we do for
phase-referenced tracks. After removing the instrumental delay, we ran
FRING in AIPS on delay calibrators to solve for the multi-band delays
and apply this correction to the target source.
    
In our 2013 and 2014 data, we noticed a fast fringe phase drifting
over time and lower fringe amplitude on all GBT baselines on
calibrator scans, after all calibrations were applied. To quantify
this issue, we fit the fringe phase, delay and rate using the
strongest maser lines in NGC 5765b (at velocity between 8294 and 8301
km~s$^{-1}$, with flux $>$ 100 mJy) on the GBT-VLA baseline with an
integration time of 30s. Then we smoothed the fitted results with a
boxcar window of 5 minutes and got a residual fringe rate (RFR) of
$\sim$ 4 mHz on the GBT over the whole observation. We repeated this
measurement in all our VLBI datasets and found this RFR shows up on
all calibrator scans and target scans, and it remains constant over
the entire track and also between tracks. Finally we corrected the RFR
using AIPS task CLCOR by adding in the rate correction before
splitting the data for self-calibration.

After calibrating each of the tracks and doing preliminary imaging, we
identified the best tracks and combined them in the following way: For
data observed in early 2013, we combined data from BB321C, BB321E1,
BB321F and BB321N in UV space, as they were observed within 2 months
of each other. Likewise, for data observed in late 2013 until early
2014, we combined BB321T, BB321Y2, BB321Y3 and BB321Y4 in UV
space. Then we combined these two maps in the image plane by applying
a weighted average to their positions for each velocity channel, with
each velocity channel matched within 0.3 km~s$^{-1}$ between the two
datasets. The channel size of the final datasets is $\sim$ 1.7
km~s$^{-1}$. The two maps were aligned by referencing to the position
of the strongest feature in the systemic part of the spectrum,
assuming they are the same feature with no position change (if we
assume a Keplerian rotation velocity of 600 km~s$^{-1}$ for the
strongest feature, the sky position change of this feature over 1 yr
is only $\sim$ 0.001 mas). This gives the final VLBI map we use in the
following analysis, as shown in Fig.~\ref{figure:vlbi-map}.

\subsection{VLBI observation results}
In general, our VLBI map (Fig.~\ref{figure:vlbi-map}) shows a linear
configuration of all maser features along the position angle (P. A.)
of 145$^{\circ}$ (defined with respect to the red-shifted masers, as
measured from due north to the east). The position for each maser spot
is referenced to the strongest line in the systemic part of the
spectrum. By assuming a Hubble constant of 70 km~s$^{-1}$ Mpc$^{-1}$
and recession velocity of 8345 km~s$^{-1}$ for NGC 5765b, we get a
nominal distance of 119 Mpc. The high-velocity masers extend between
0.5 to 2.0 mas from the coordinate origin, which corresponds to $\sim$
0.3 to 1.2 pc. We note the blue-shifted and red-shifted features are
distributed asymmetrically with respect to the systemic features, and
the disk shows a clear warp in P.A. Maser spots on the
red-shifted side are concentrated into 5 clumps, while the
blue-shifted spots are more evenly distributed and are roughly divided
into 2 parts. We discuss such clumpiness in more detail in Section 5.
     
\begin{figure}[h]
\epsscale{1.0}
\plotone{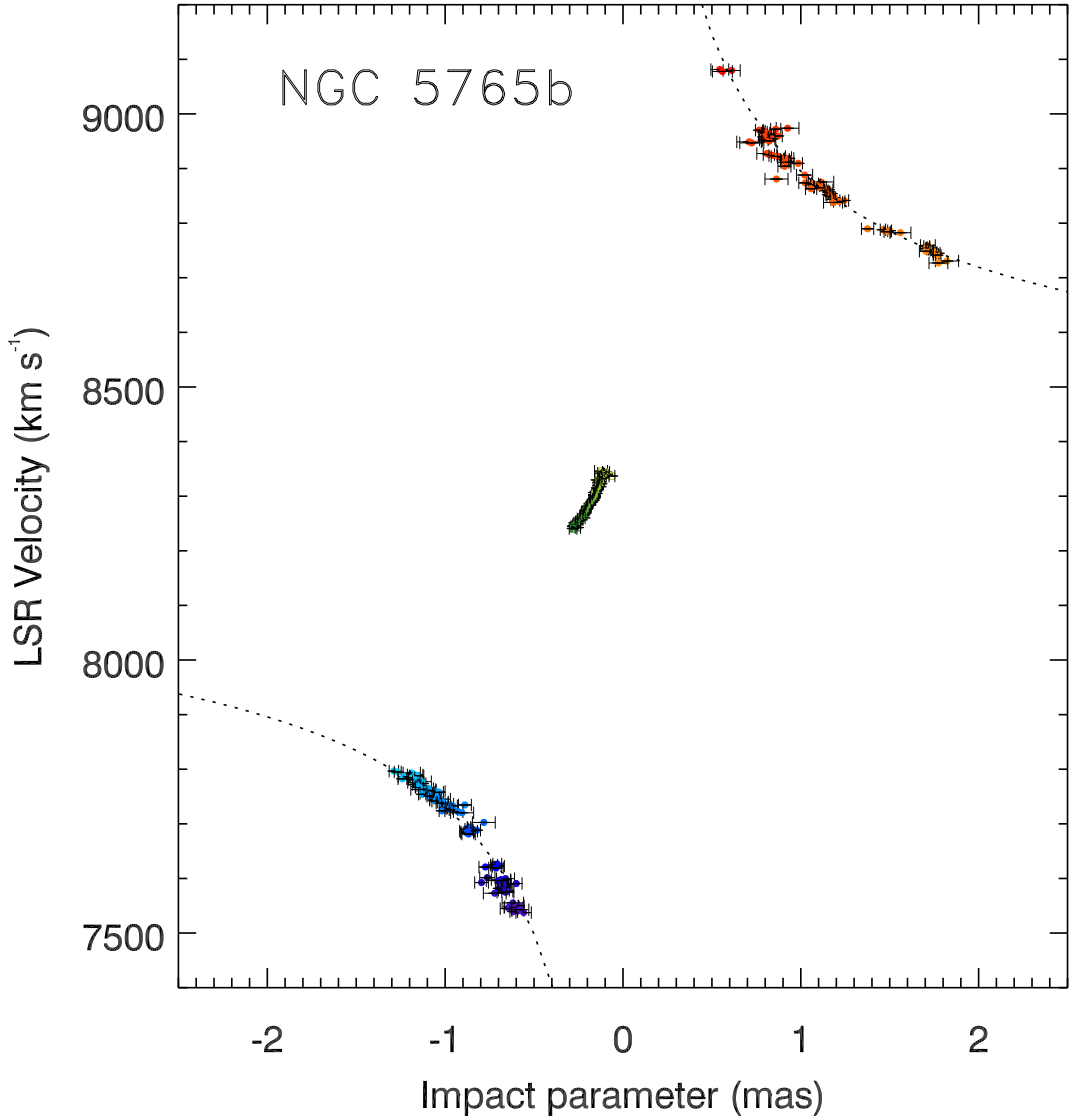} 
\caption{\footnotesize 
The position-velocity diagram from the combined VLBI results, with a flux
cutoff of 3 mJy, which is at the 10$\sigma$ level. The dotted line
shows the best-fit Keplerian rotation curve assuming an edge-on thin
disk model without disk warping. This model gives an enclosed mass of
4.4 $\pm$ 0.44 $\times10^{7}M_\odot˜$ and a recession velocity of
8304 km s$^{-1}$. The origin on the X-axis marks the position of the
dynamic center. For most of the high-velocity features, their
deviation from the fitted rotation curve is less than 5\% of their
rotation velocity.
 \label{figure:pv-map}
}
\end{figure}

The spatial distribution of maser spots in NGC 5765b is consistent
with a thin-disk model. To check the kinematics of the maser disk, we
construct the position-velocity (P-V) diagram by defining the impact
parameter as r = $\sqrt{(x-x_{0})^{2}+(y-y_{0})^{2}}$, in which
x$_{0}$ and y$_{0}$ are the VLBI position for the strongest maser spot
at 8295.91 km~s$^{-1}$. Then we fit the high-velocity data with a
Keplerian rotation curve, assuming an edge-on disk
($i_{0}$=90$^{\circ}$) with no inclination warping and also assuming that all the
high-velocity maser spots are located along the midline (defined as
the radial line through the disk perpendicular to the line of
sight). The best fit yields an enclosed mass of 4.4 $\pm$ 0.44
$\times$ 10$^{7}$M$_\odot˜$, with the fitted recession velocity
8304 km~s$^{-1}$ and a position offset of the dynamic center of 0.011
mas in R.A. and -0.162 mas in DEC from the coordinate origin. In
Fig.~\ref{figure:pv-map} we show the P-V diagram, which is centered on the
fitted dynamic center. The best-fit Keplerian curve is overplotted as
the dotted lines. Most of the high-velocity features deviate from the
best-fit Keplerian curve by less than 5\% of their rotation
velocity. We treat this as the main contributor to the uncertainty on
the black hole mass measurement since this simple model doesn't include fitting
the inclination warp of the disk. So the final uncertainty for the
black hole mass is about 10\%.
    
While this simple model does not account for any inclination warping on
the maser disk, the VLBI map and P-V diagram together still confirm
that masers in NGC 5765b trace a thin, sub-parsec Keplerian disk. We
treat these results as an initial guess for the central black
hole mass, recession velocity and position of the dynamic center in
the more detailed modeling of the maser disk (Section 4).

\section{ACCELERATION MEASUREMENTS}

\subsection{Monitoring observations}

To measure the secular drift of maser velocities in NGC 5765b over
time, we obtained monthly GBT monitoring observations.  Each observation lasted
$\sim$ 3 hours, from which we reach an rms noise of $\sim$ 1.2 mJy, at
a channel size of 24.4 kHz with Hanning smoothing. All together we
obtained 18 epochs of observations from 2012 February to 2014 May,
which bracket our VLBI observations. We list the details on the
monitoring observations in Table~\ref{table:GBT observation}. We note
that because of a scheduling issue, we didn't get monitoring
observations in 2014 January. Since our 2014 VLBI observation took
place at this time, we used the VLBI spectrum to fill in that
epoch. In this regard we have 19 epochs of data used in the
acceleration measurements.
    
\subsection{Acceleration of the systemic masers}
\subsubsection{Measuring ``by eye"}
	
\begin{figure}[h]
\epsscale{1.00}
\plotone{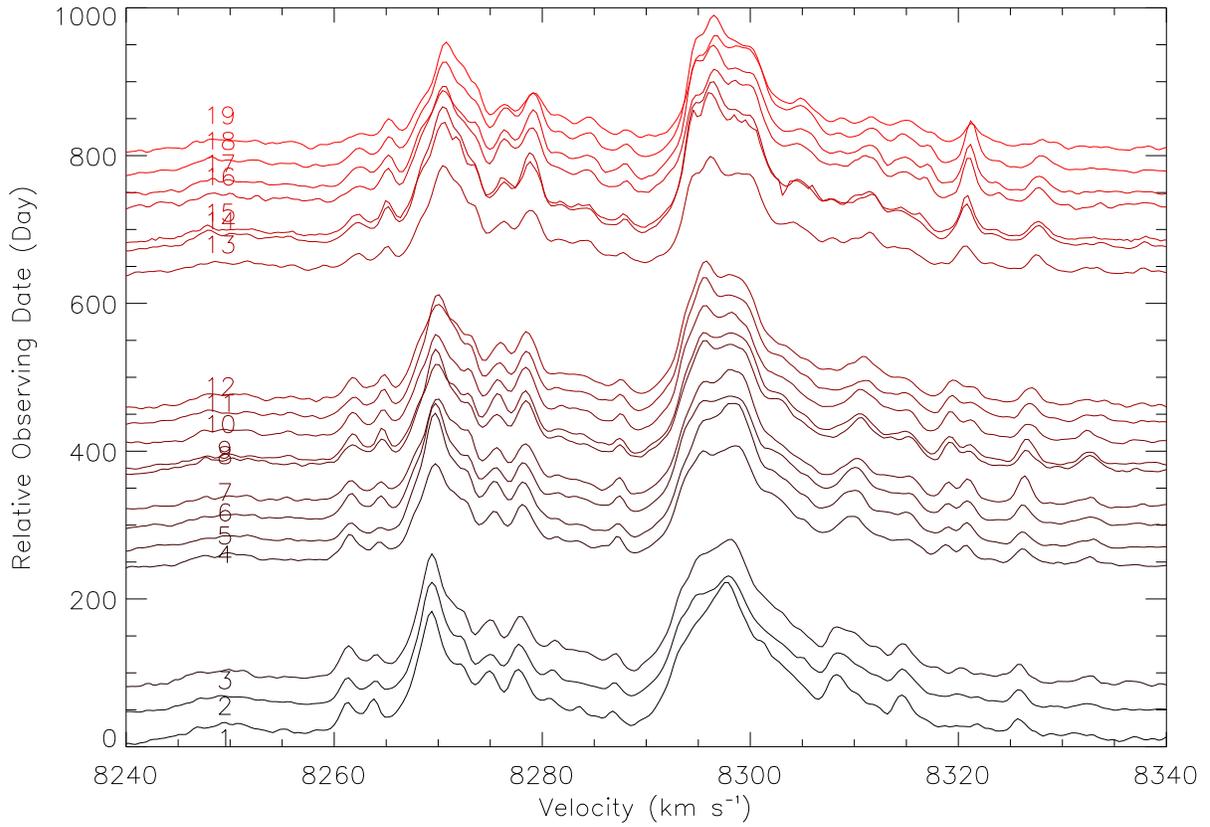} 
\caption{\footnotesize 
Stacked spectra from GBT monitoring data, zoomed on the systemic
  part, with the date referenced to 2012 February 26. Data from epoch 15
  comes from VLBI tracks BB321T Y2 Y3 and Y4 combined, as a complement
  to the unscheduled GBT monitoring observation.
 \label{figure:stack-sys}
}
\end{figure}

\begin{figure}[H]\centering
\includegraphics[width=.25\textwidth]{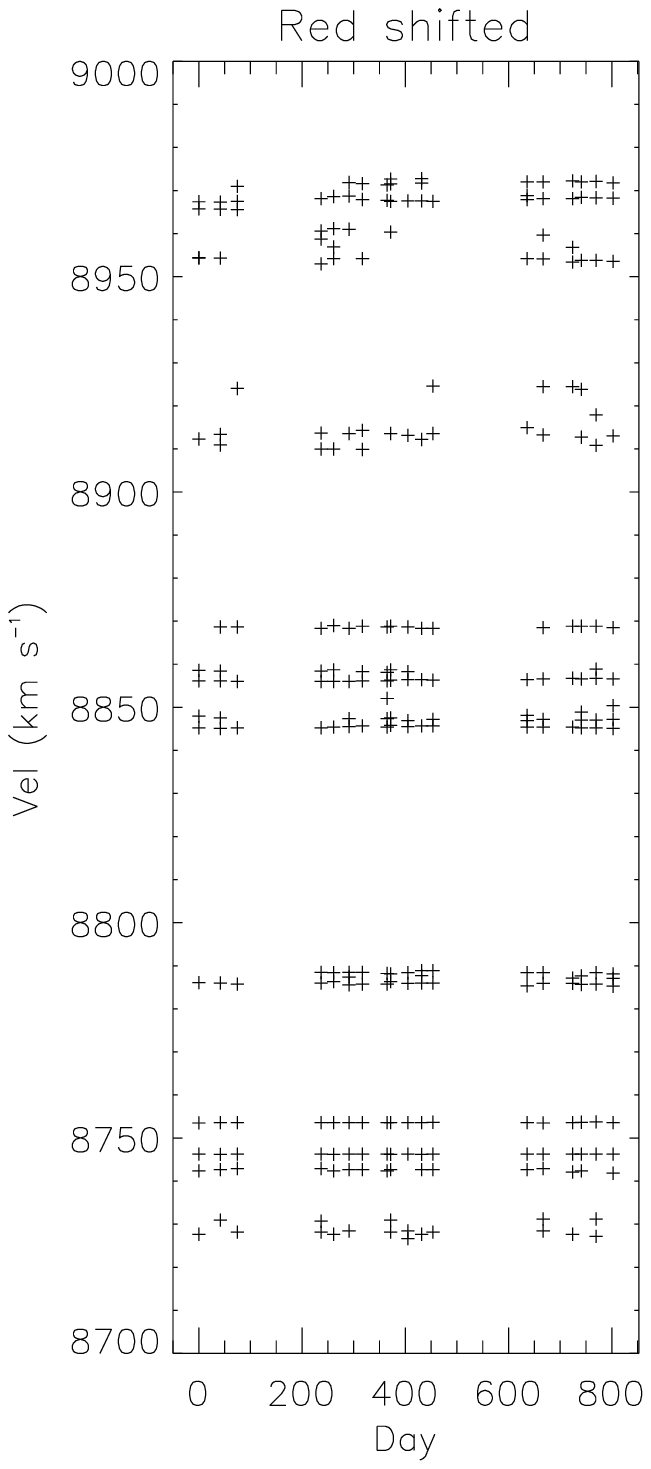}
\includegraphics[width=.25\textwidth]{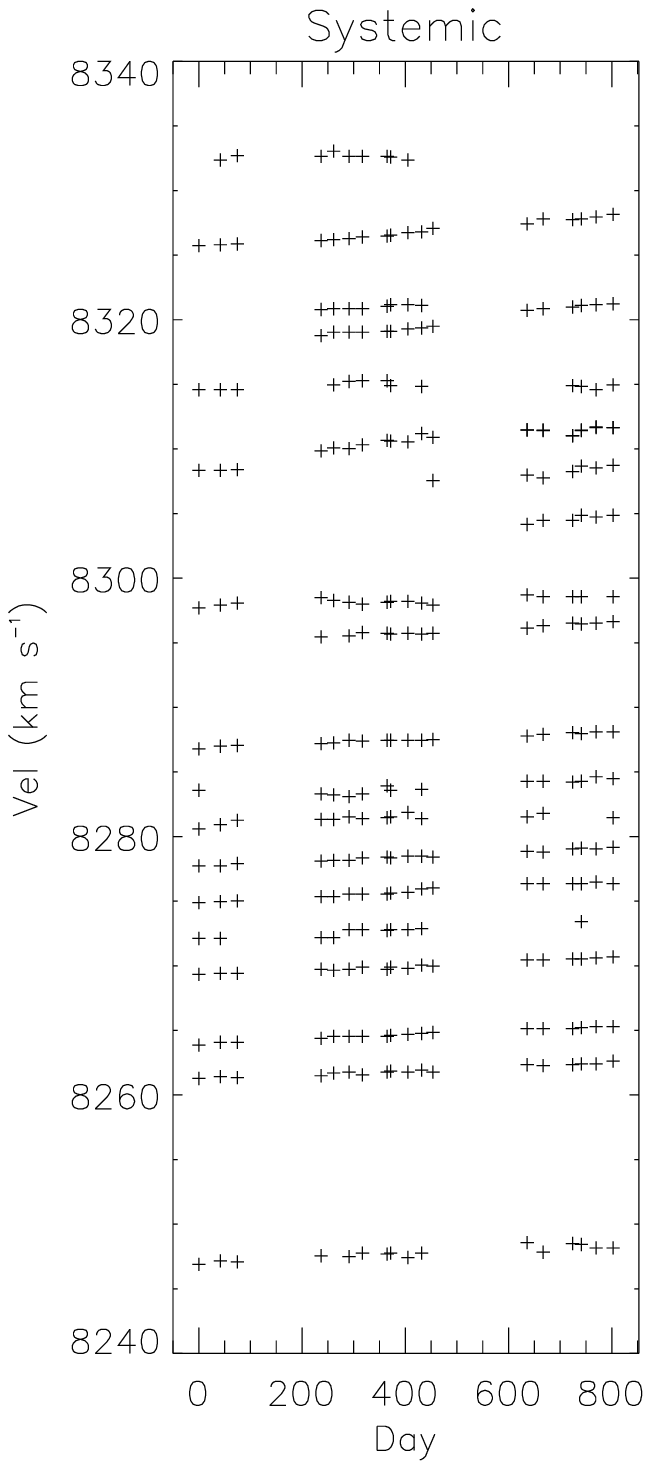}
\includegraphics[width=.25\textwidth]{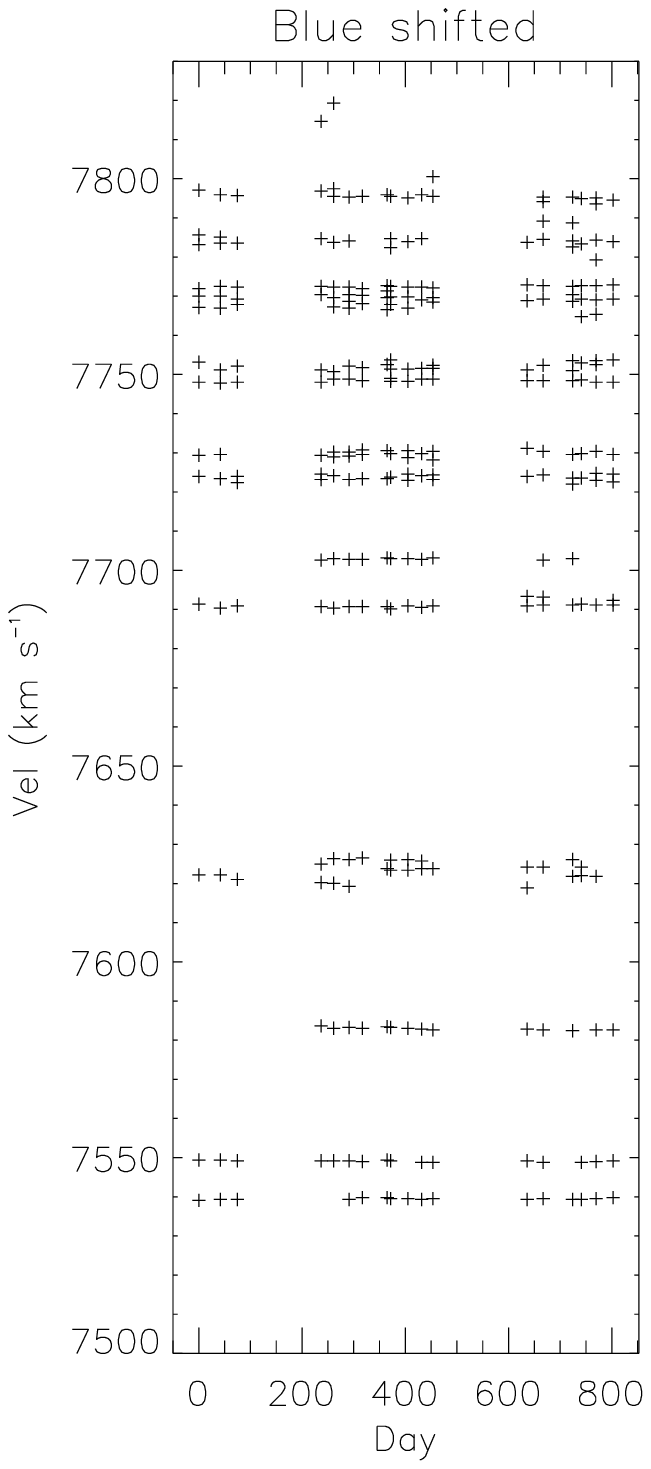}
\caption{\footnotesize 
These acceleration plots show velocities of the maser line peaks in both the systemic and
high-velocity parts. Each cross marks a peak velocity at the given epoch,
identified by eye.  The date on the X-axis
is referenced to the first GBT monitoring observation data, 
2012 February 26. The ``negative'' slope seen in several systemic
features (\eg $\sim$ 8300 km s$^{-1}$ in the middle epochs) appears
to result from lines blending together and the emerging of new 
peaks at adjacent velocities. This is better
illustrated in Fig.~\ref{figure:stack-sys}.
 \label{figure:clicking}
}
\end{figure}

We show all 19 epochs of monitoring spectra in
Fig.~\ref{figure:stack-sys}, from which we can see the overall trend
for line profile drifting and trace each individual component. As a
first approximation, we use the peak of each distinct maser feature to
trace the underlying acceleration. Those peaks were identified by eye
and the results plotted in Fig.~\ref{figure:clicking}. From this, we
get an initial estimate of the number of Gaussian components
associated with the spectrum and the magnitude of the
accelerations. From Fig.~\ref{figure:clicking} it is clear that the
systemic features can be traced easily over the two year monitoring
period, and show distinct accelerations. The features between 8260
km~s$^{-1}$ and 8290 km~s$^{-1}$ all have similar accelerations of
$\sim$ 0.6 km~s$^{-1}$ yr$^{-1}$. Features outside that velocity range
show more varied accelerations, between 0.5 and 1.2 km~s$^{-1}$
yr$^{-1}$.
	
The ``by eye" method does not adequately distinguish blended
features. We also note that some systemic features (\eg $\sim$ 8300 km
s$^{-1}$ in the middle epochs) show a ``negative'' slope, which we
think is not a sign of ``negative'' acceleration, but may have
resulted from several lines blending together and the emerging of new
peaks at different velocities, as the overall spectral profile shows a
clear change (which is better illustrated in
Fig.~\ref{figure:stack-sys}).
	
So we get initial estimates of line accelerations using the ``by eye''
method. We then use these estimates (only the positive ones) as the
initial values in a least squares determination of accelerations, as
described in the next section.
 
\subsubsection{Method 1}
	
To determine the best fit accelerations for each maser component, we
decompose the overall profile into a series of Gaussian components and
measure the drift of each component as the acceleration, using a
least squares fitting algorithm. This is the so-called ``Method 1" in
previous MCP papers.
	
We follow the fitting process described in Reid et al. (2013), in
which for the first step, we specify the velocity and acceleration for
each component at a reference epoch, keep the line width fixed to 2
km~s$^{-1}$ (the typical line widths for the systemic masers in NGC
5765b are between 2 to 3 km~s$^{-1}$, and here we use 2 km~s$^{-1}$ as an
initial guess), and let the program set the amplitude of each
component at the corresponding velocity from the real data. After this
step, we have an initial mask of the Gaussian decomposition. We
minimize the residual $\chi^{2}$ between our Gaussian decomposition
model and the real data by doing the fitting iteratively. In the
second step, we fit the amplitude, width and velocity together for
each component, but keep the acceleration unchanged. Finally we
include the acceleration in the fitting and stop when the residual
$\chi^{2}$ reaches a stable minimum.
	
A visual inspection shows the systemic maser spectrum divides mainly
into two parts around 8290 km~s$^{-1}$, with the features between 8260
km~s$^{-1}$ to 8290 km~s$^{-1}$ (named ``clump 1") more distinct,
while features between 8290 km~s$^{-1}$ to 8340 km~s$^{-1}$ (named
``clump 2") more blended, especially around 8295 km~s$^{-1}$.  So we
fit the accelerations in the two clumps separately, with the 
procedure mentioned above. The final reduced $\chi^{2}$ for the two
clumps are 1.673 and 1.758, respectively. We note that since we have
used Hanning smoothing for our data, the optimal reduced chi-squared
value will not be unity.
	
We list the acceleration results from method 1 in
Table~\ref{table:accer-result-method1} below.
	
\subsubsection{Method 2}

\begin{figure}[h]\centering
\includegraphics[width=0.450\textwidth]{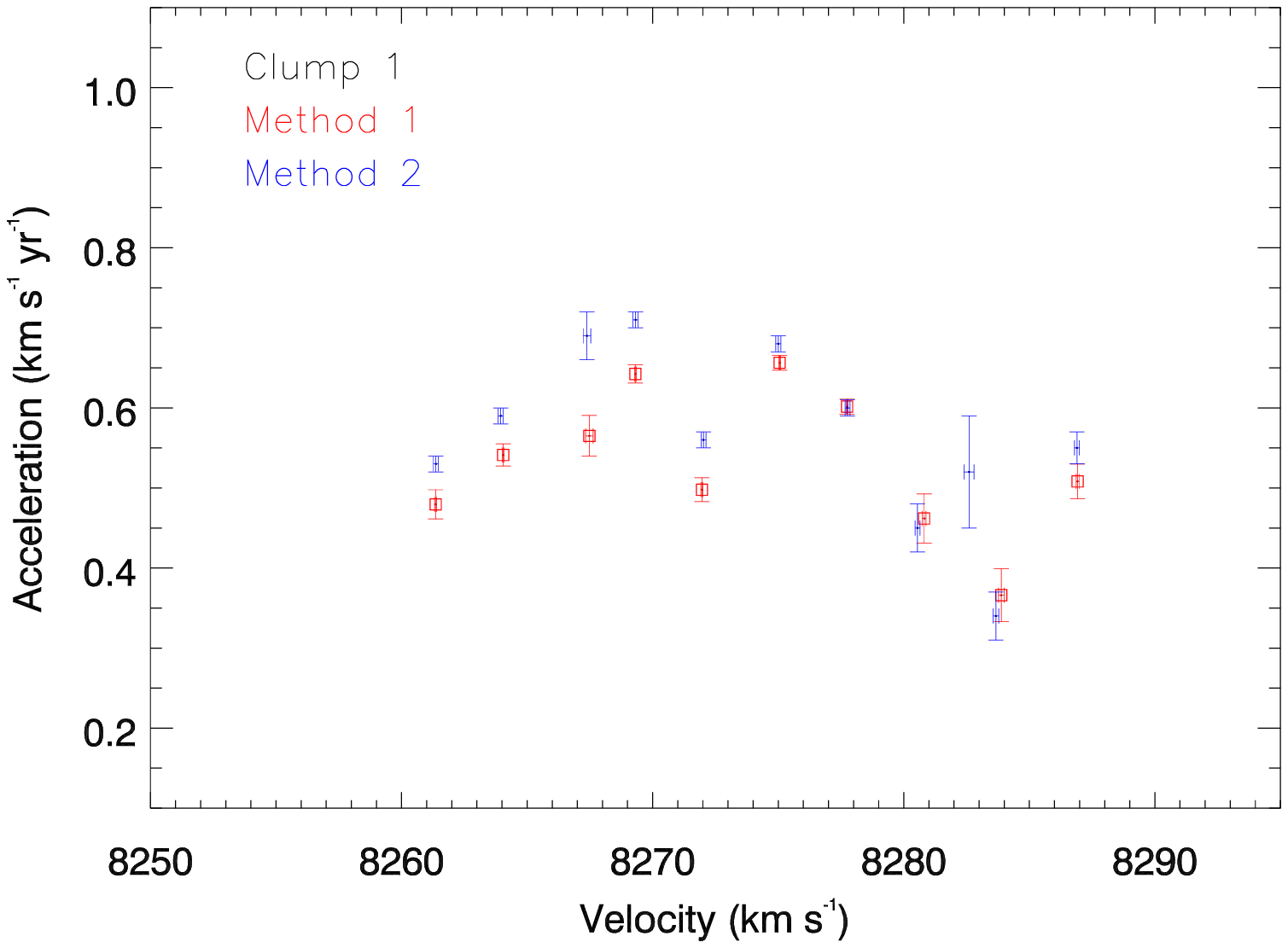}
\includegraphics[width=0.450\textwidth]{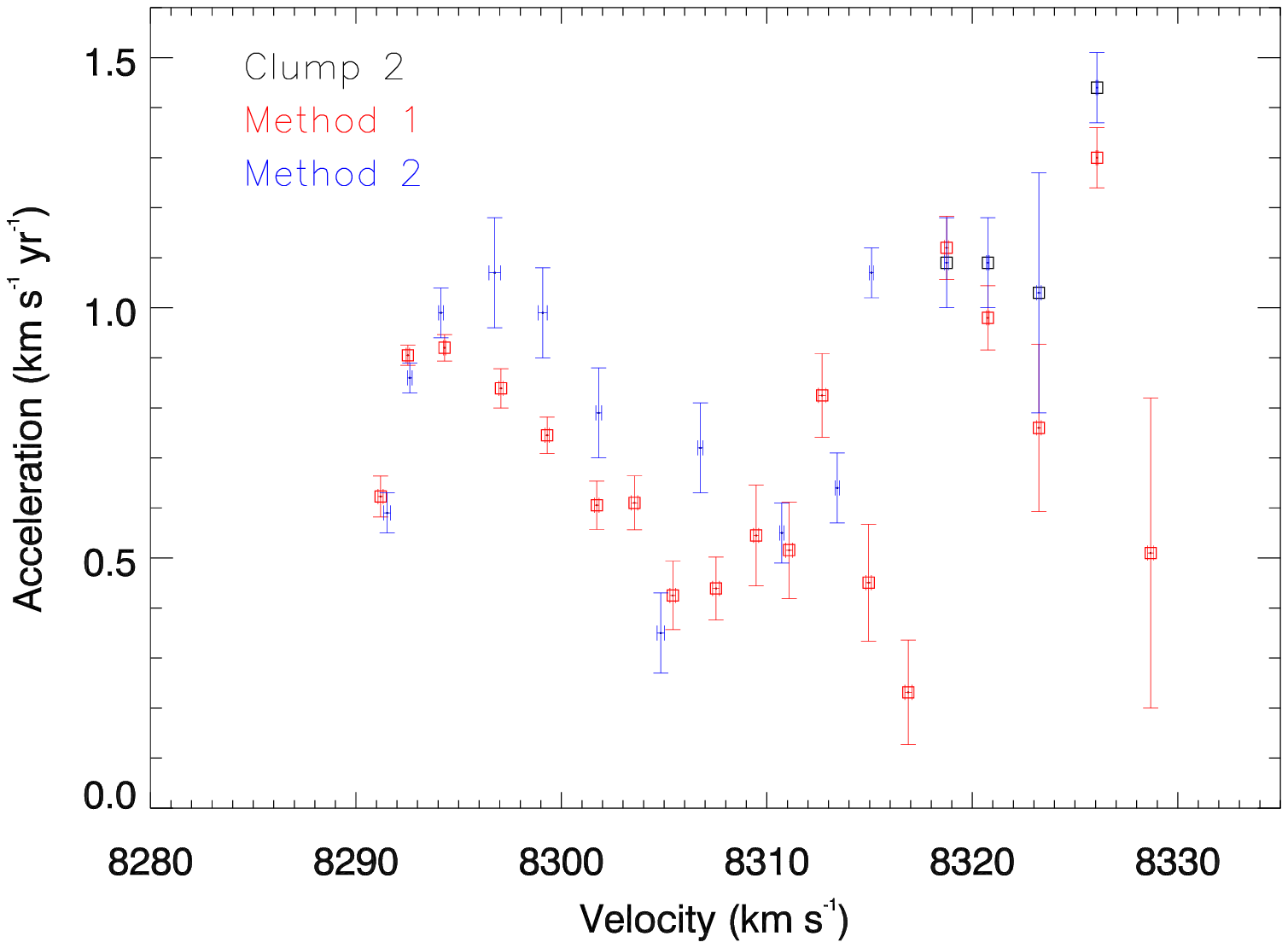}
\caption{\footnotesize 
Comparison of the acceleration results between method 1 and method 2,
for both clump 1 and clump 2 in the systemic part.
 \label{figure:accer-compare}
}
\end{figure}

To check whether the results from Method 1 depend on the initial
velocity and acceleration values, here we fit the accelerations using
the more automated Method 2 (see more description in Reid et
al. (2013) ) in which the algorithm explores a wider range of initial
parameters. Instead of specifying the number of components and the
initial guess of velocity/width/acceleration for each component, here
we only specify the velocity range for the fitting, the average width
for each component and the average separation in velocity between each
component. The program will generate multiple trials and do the
fitting separately, with each trial using a set of initial values
generated randomly for each of the components. We run the program with
100 trials to explore different starting points and keep the few best
solutions with small $\chi^{2}$ and consistency with the results
measured by eye. We show the fitting results in
Fig.~\ref{figure:accer-compare}.
        
\subsection{Error floor for accelerations of systemic masers}
	
From previous MCP experience, to account for more realistic
uncertainties in cases where the formal fitting uncertainty from
method 1 is unrealistically small (\eg less than 0.05 km~s$^{-1}$
yr$^{-1}$ in the case of NGC 5765b), we use an ``error floor" on the
acceleration result and add that in quadrature to the formal fitting
error to get the final uncertainty for each measured
acceleration. However, a single value for the ``error floor" might not
be suitable for all maser features in the case of NGC 5765b. It is
clearly shown in both Fig.~\ref{figure:stack-sys} and
Table~\ref{table:accer-result-method1} that the spectral features in
clump 1 are more distinct and accurately measured, while those in
clump 2 are more blended and show bigger scatter. Since line blending
is an important contributor to the uncertainties, a large error floor
here which may accurately represent the errors in clump 2 would
down-weight results from clump 1, while a small error floor will
overweight results from clump 2 in the other way. So for this galaxy
we use the difference between results from method 1 and method 2 for
each measured maser spot as an estimate of the error floor for each
component.
		
We plot the acceleration fitting results from both methods in
Fig.~\ref{figure:accer-compare} for comparison.

\subsection{Accelerations of the high-velocity masers}
        
The high velocity features are not as highly blended as the
systemic ones.  Considering their relative weakness and their short
lifetime (some of the blue-shifted features are only evident
for 2 to 3 epochs), we measured
their accelerations using only the ``by eye'' method. We list these results in
Table~\ref{table:accer-result-highvelo}.
	
Besides the formal fitting error, we also calculate the rms scatter in
both the red-shifted and blue-shifted results, which are 0.155
km~s$^{-1}$ yr$^{-1}$ (red) and 0.275 km~s$^{-1}$ yr$^{-1}$ (blue) and
0.200 km~s$^{-1}$ yr$^{-1}$ (combined).  Overall, most of the measured
results are within this rms level, with only maser features at 8856.32
km~s$^{-1}$ and 7582.95 km~s$^{-1}$ showing clear offsets from zero
acceleration. For a high-velocity feature with rotation velocity of
600 km~s$^{-1}$ in NGC 5765b, an acceleration of 0.2 km~s$^{-1}$
yr$^{-1}$ along the LOS direction requires a 16$^{\circ}$ deviation
from the midline on the disk. Thus we conclude that most of the high
velocity features locate on the mid-line (the diameter through the
disk perpendicular to the line of sight) of the maser disk with
deviations less than $\sim$ 16$^{\circ}$, with few exceptions.

\section{DETERMINATION OF $\Ho$ }

\subsection{Model fitting of the accretion disk}

To reconstruct the geometry and spatial distribution of maser spots
from the observed data, here we fit a warped disk model to the VLBI
and acceleration data, using a Bayesian approach. This model can
accommodate 14 global parameters describing the disk geometry,
including the Hubble constant $\Ho$, the central black hole mass M,
position of the dynamic center on the sky $X_{0}~\&~Y_{0}$, recession
velocity and peculiar velocity of the galaxy $V_{0}~\&~V_{p}$,
position angle and first/second order of warping on the P.A. direction
$PA, \partial p/\partial r, \partial^2 p/\partial r^2$, inclination
angle and first/second order of warping on the inclination direction
$i, \partial i/\partial r, \partial^2 i/\partial r^2$ and two
additional parameters describing the eccentricity of the disk e and
the change of the azimuth angle for the major axis of the elliptical
with radius $\partial p_{az}/\partial r$. Apart from these global
parameters, there are also two parameters (r and $\phi$) describing
the coordinates on the disk for each maser spot. The input data for
the model consists of the sky position, velocity and acceleration for
each maser spot. The program uses a Markov chain Monte-Carlo (McMC)
approach to sample the parameter space and applies the
Metropolis-Hastings algorithm to explore the probability distribution
function for each of the global parameters. To better sample parameter
space, the program allows multiple ``strands", defined as independent
runs of the McMC with different starting conditions, to run
simultaneously. A more detailed description about the disk model and
fitting process can be found in Reid et al. (2013) and Humphreys et
al. (2013).

All together there are 212 maser spots mapped for NGC 5765b. We list
the input data in Table~\ref{table:disk fitting data}. We estimated
the initial value for the position angle, black hole mass, recession
velocity and black hole position from our fitting of the P-V diagram
in Section 2, while the initial value for $\Ho$ in each strand was
evenly selected between 60 and 80 km~s$^{-1}$ Mpc$^{-1}$. Except for recession velocity and
peculiar velocity, we didn't include a prior uncertainty in any
other global parameters as we try to keep our fitting result away from
any existing bias. We also note that even though the initial value for
$\Ho$ was chosen between 60 and 80 km~s$^{-1}$ Mpc$^{-1}$, $\Ho$ could vary to any number as
there are no constraints on it during each trial of fitting.

NGC 5765b sits in a galaxy pair together with NGC 5765a, and the
recession velocity for each of them are (a) 8469 km~s$^{-1}$ and (b)
8345 km~s$^{-1}$. Assuming the dynamic center of this galaxy pair has
the averaged recession velocity of 8401 km~s$^{-1}$, then the
additional uncertainty in peculiar velocity caused by this galaxy pair
is $\sim$ 70 km~s$^{-1}$, which is less than 1\% of the recession
velocity of NGC 5765b. Apart from this, we don't have any constraints
on whether NGC 5765b deviates from the Hubble flow. Nevertheless, we
include the peculiar velocity in our fitting so that the uncertainty
of peculiar velocity contributes appropriately to our final
uncertainty in $\Ho$. As a conservative treatment, we use a peculiar
velocity of zero with prior uncertainty of 250 km~s$^{-1}$ in our
final fitting.

\begin{figure}[h]
\epsscale{1.0}
\plotone{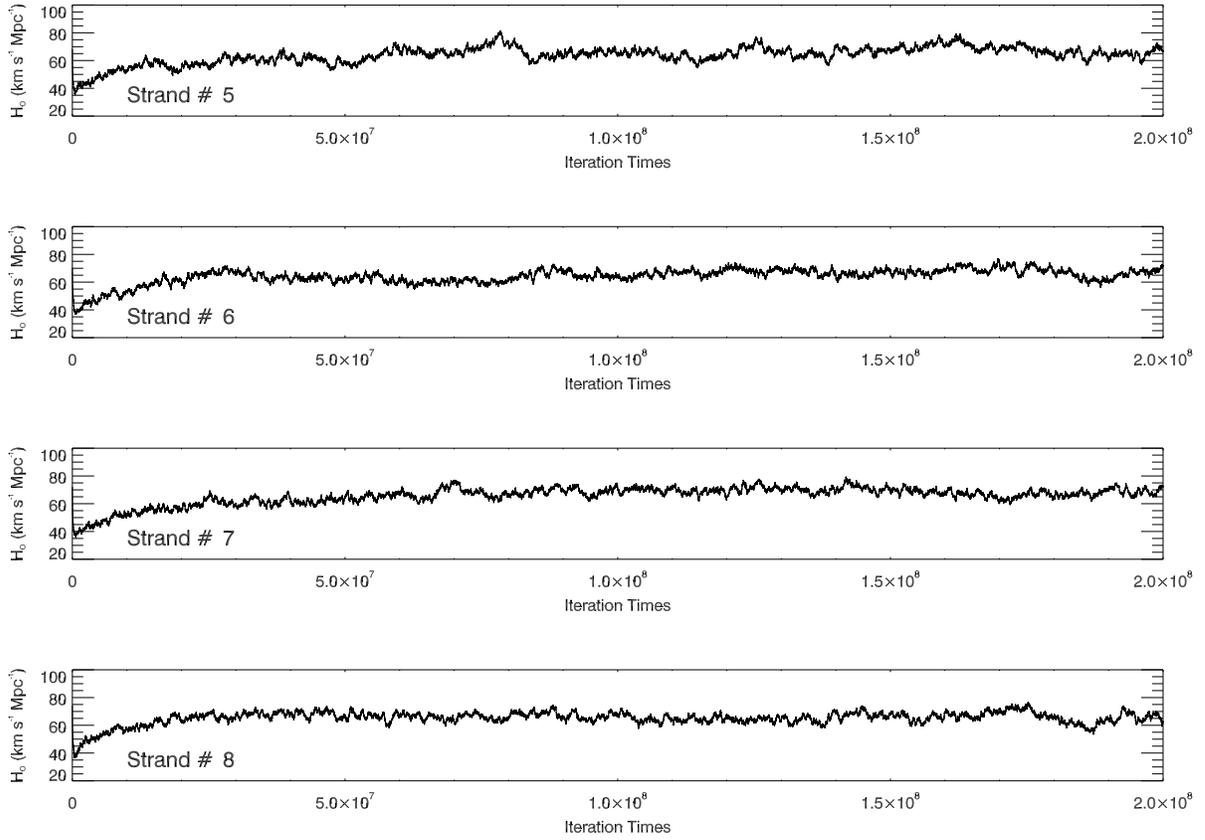} 
\caption{\footnotesize 
Example of the $\Ho$ versus iteration number plot for several strands
in our fit. Note that $\Ho$ starts from an initial value between
60 and 80, then quickly drops to $\sim$ 40 at the very beginning
iterations, then gradually converges to the final result. In our
calculation of the posteriori probability distribution function, we
discard records from the first 20\% of the iterations to avoid
unconverged trials in our final result.
 \label{figure:H0iter}
}
\end{figure}

\begin{figure}[h]
\epsscale{1.0}
\plotone{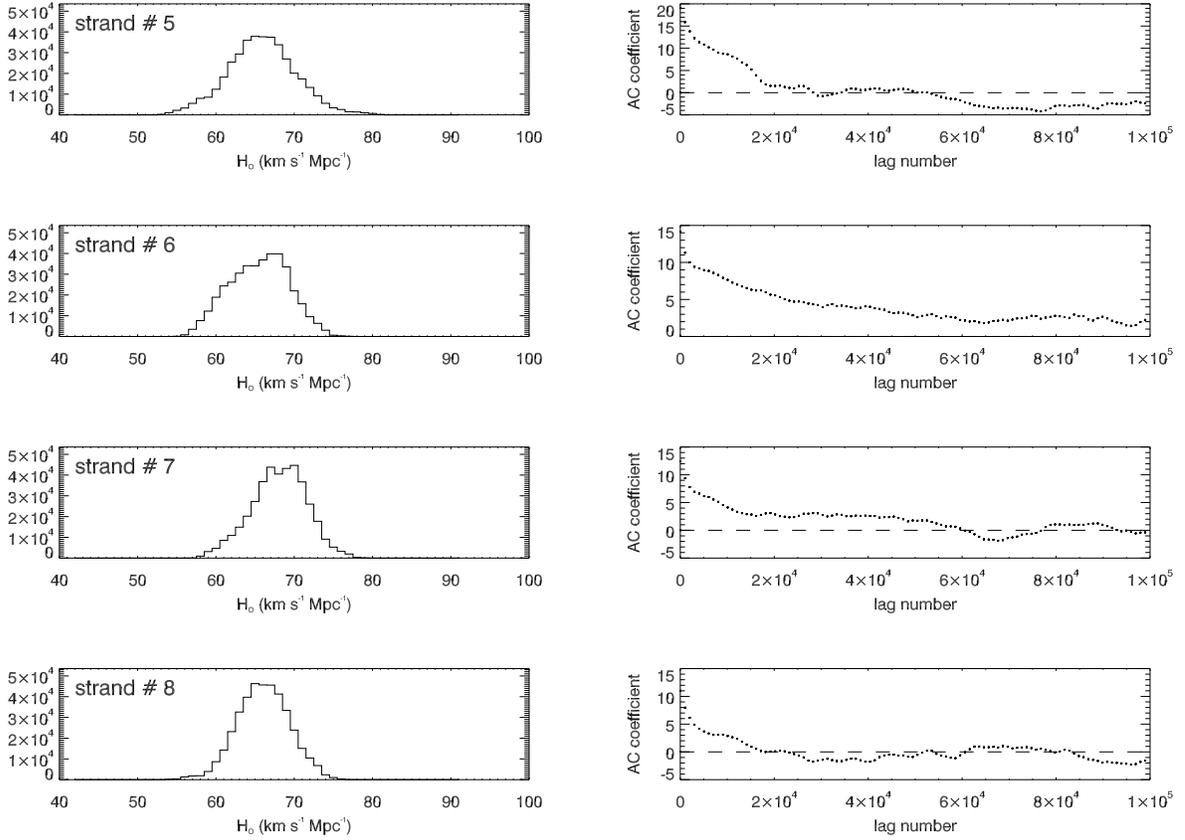} 
\caption{\footnotesize 
Example of the probability distribution of $\Ho$ and the
auto-correlation (AC) function versus lag number for several
strands. The AC functions are used to check for fitting
convergence. We require the AC function to drop to near zero before
40\% of the maximum lag number. The maximum lag number is defined as
25\% of the total number of recorded McMC trials in each strand. In our
case, we record every 400th record of the total $2\times10^{8}$
trials, after discarding the first 20\%, so the maximum lag number is
$1\times10^{5}$. Here strand 6 is not quite converged according to our
criteria. Usually if 6 or 7 out of the total 10 strands are converged,
we consider the total fitting result as converged.
 \label{figure:AC}
}
\end{figure}

In our final model fitting we use 10 global parameters, where we have
excluded the second order of warping in both the P.A. and inclination 
directions, and the two parameters for eccentricity, as they were 
insignificant to the
final result. We use 10 strands in our fitting, with each strand
containing $2\times10^{8}$ trials, which is sufficient to reach
convergence. The convergence is indicated by the autocorrelation (AC)
function for each model parameter.  We require the AC
function to drop near zero before 40\% of the largest lag number. The
largest lag number is defined as 25\% of the maximum number of McMC
trials that have been recorded for each strand. Usually if 6 or 7 out
of the total 10 strands have converged, we consider the total fitting
result as converged. We show an example plot of the $\Ho$ versus trial
iteration number for several strands in
Fig.~\ref{figure:H0iter}. Another example of the AC function with lag
number and also the $\Ho$ probability distribution for the same
strands are shown in Fig.~\ref{figure:AC}.

\begin{figure}[h]\centering
\includegraphics[width=.32\textwidth]{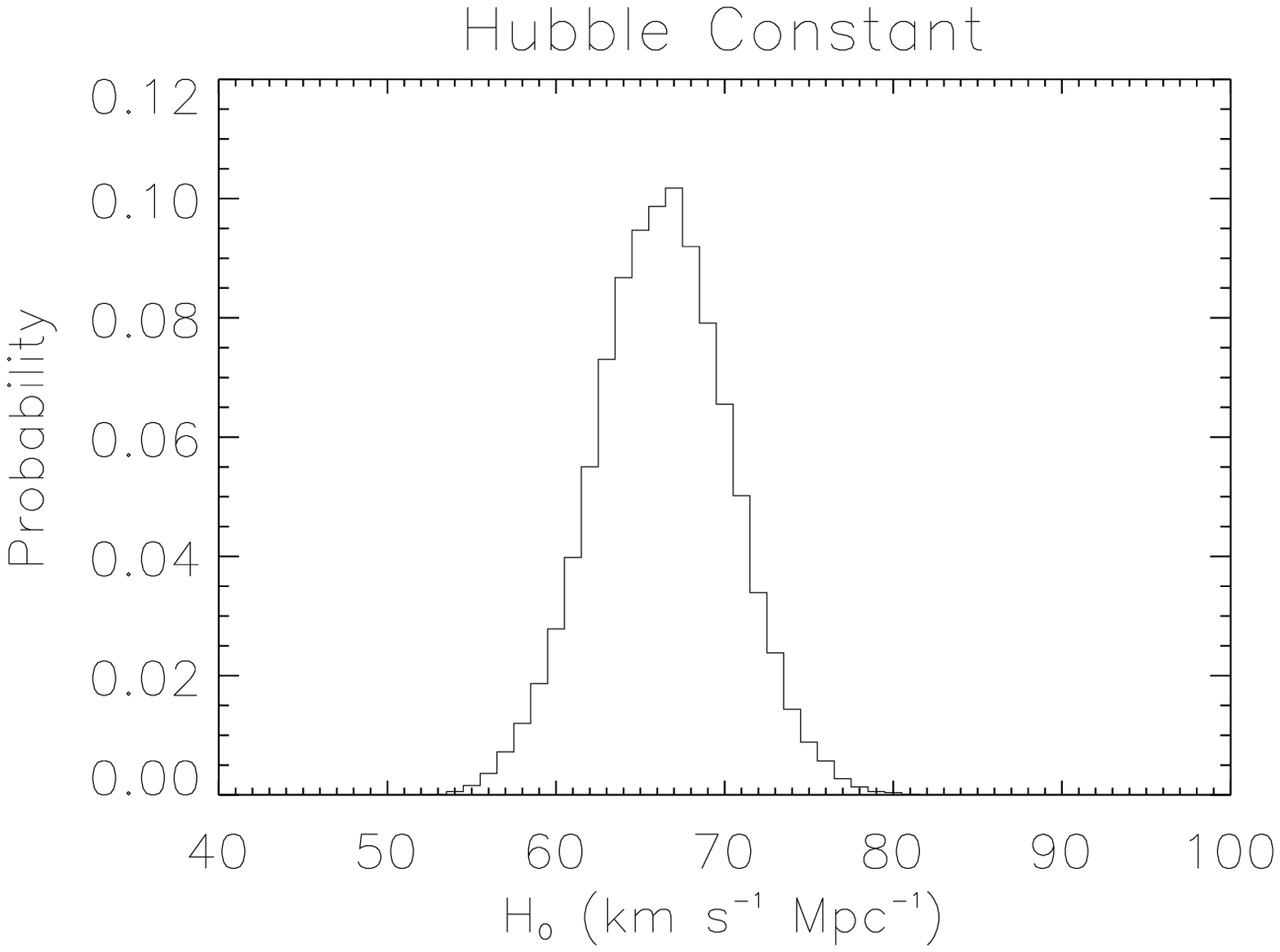}
\includegraphics[width=.32\textwidth]{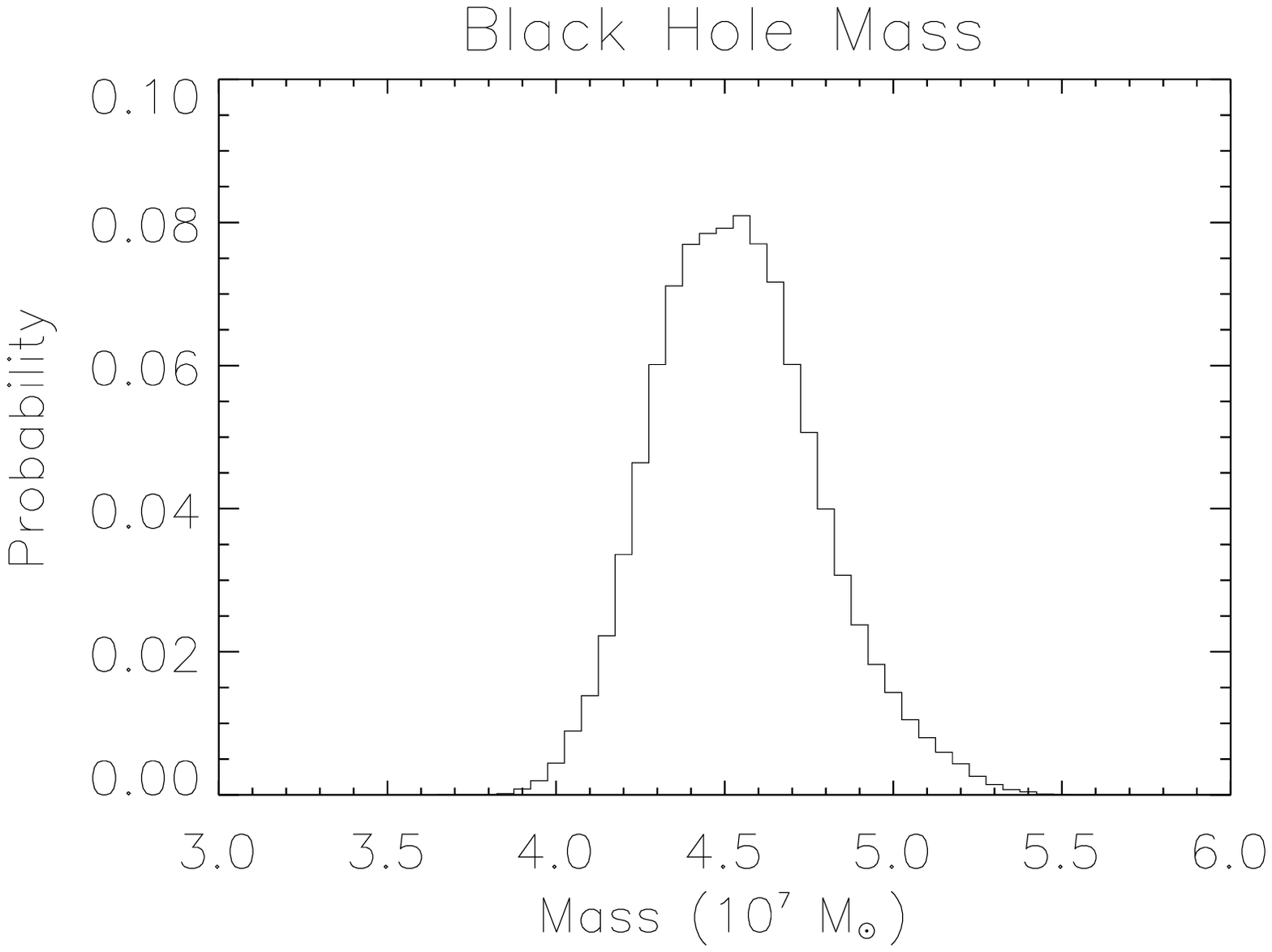}
\includegraphics[width=.32\textwidth]{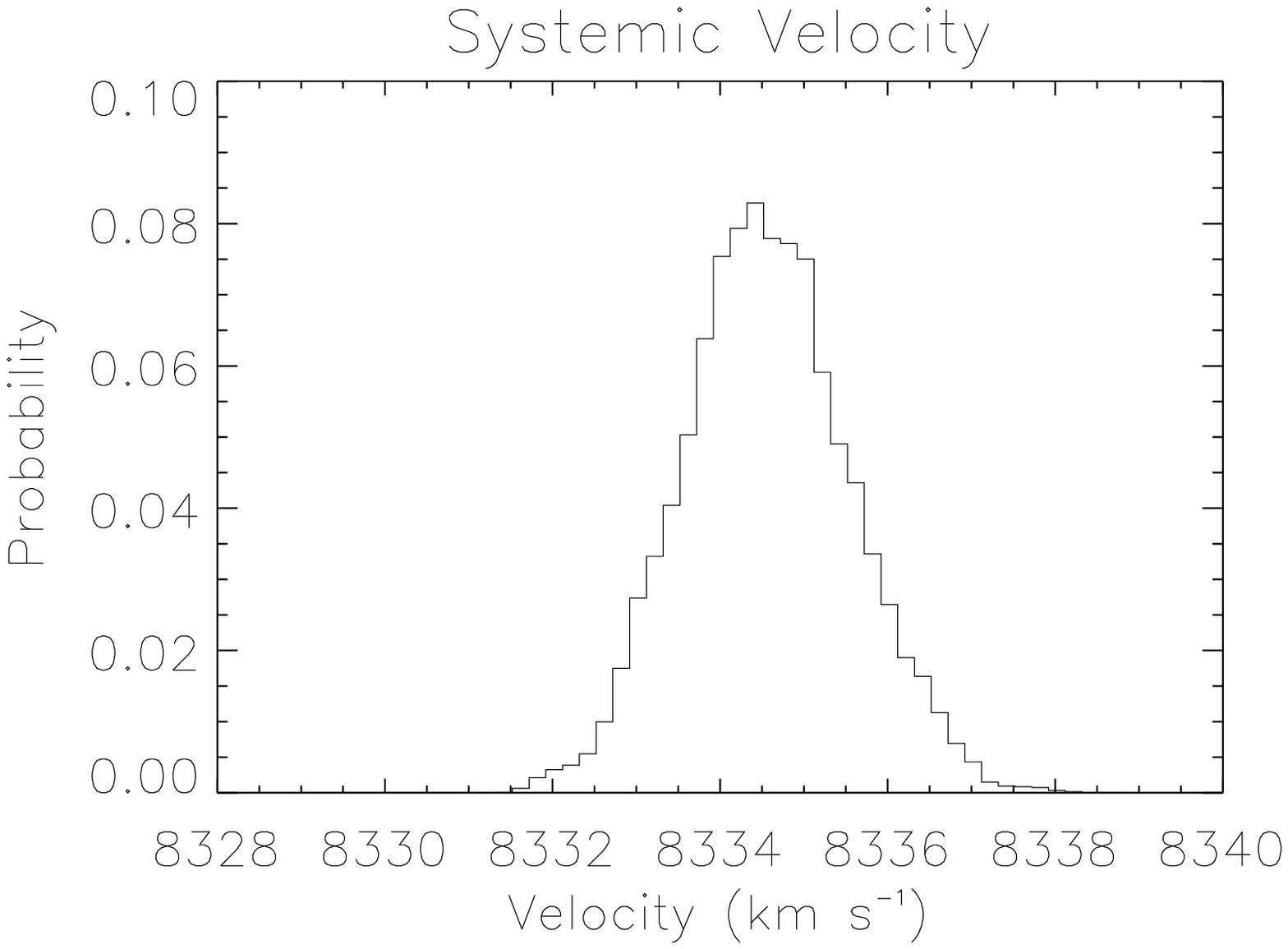}
\includegraphics[width=.32\textwidth]{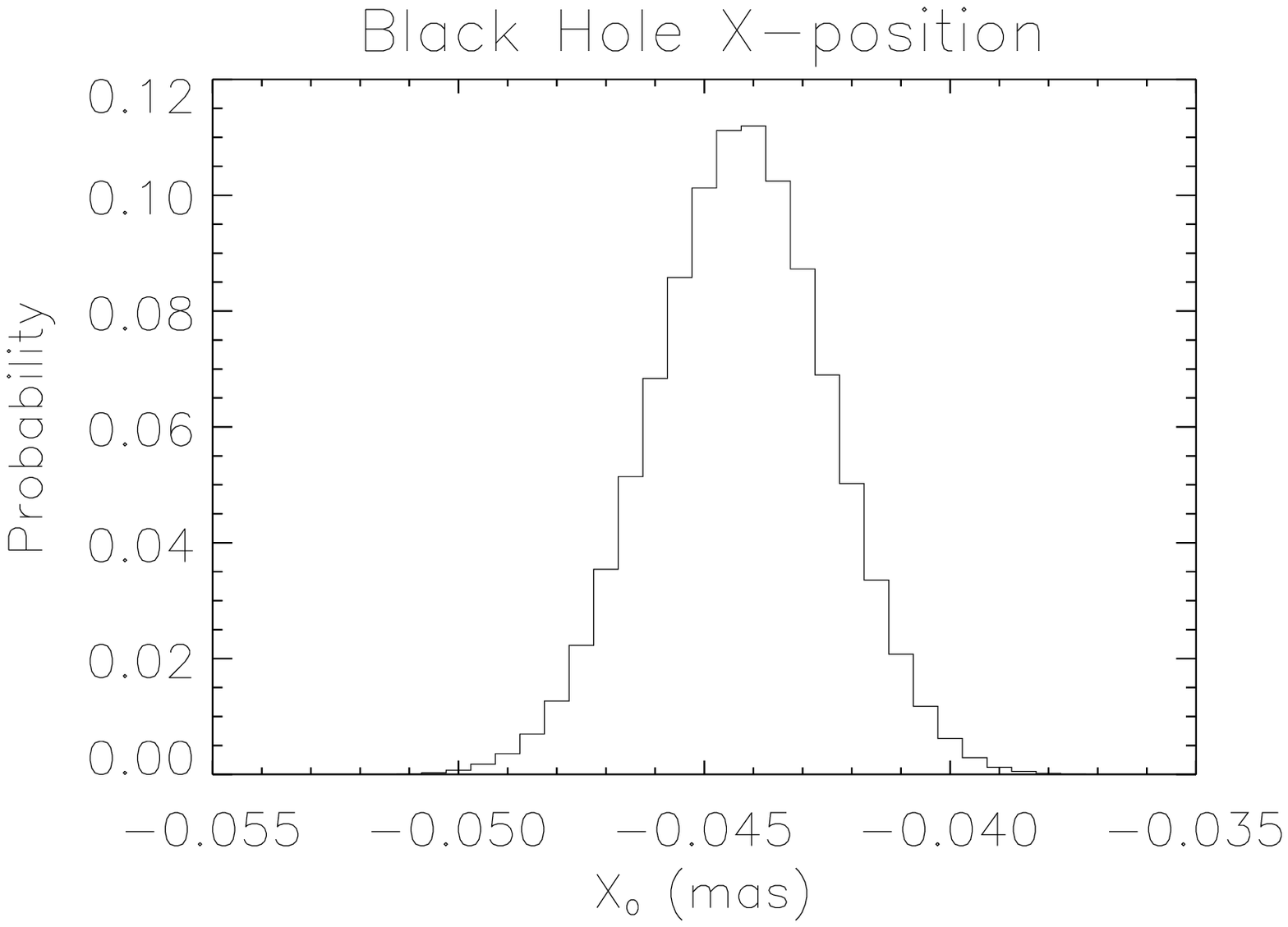}
\includegraphics[width=.32\textwidth]{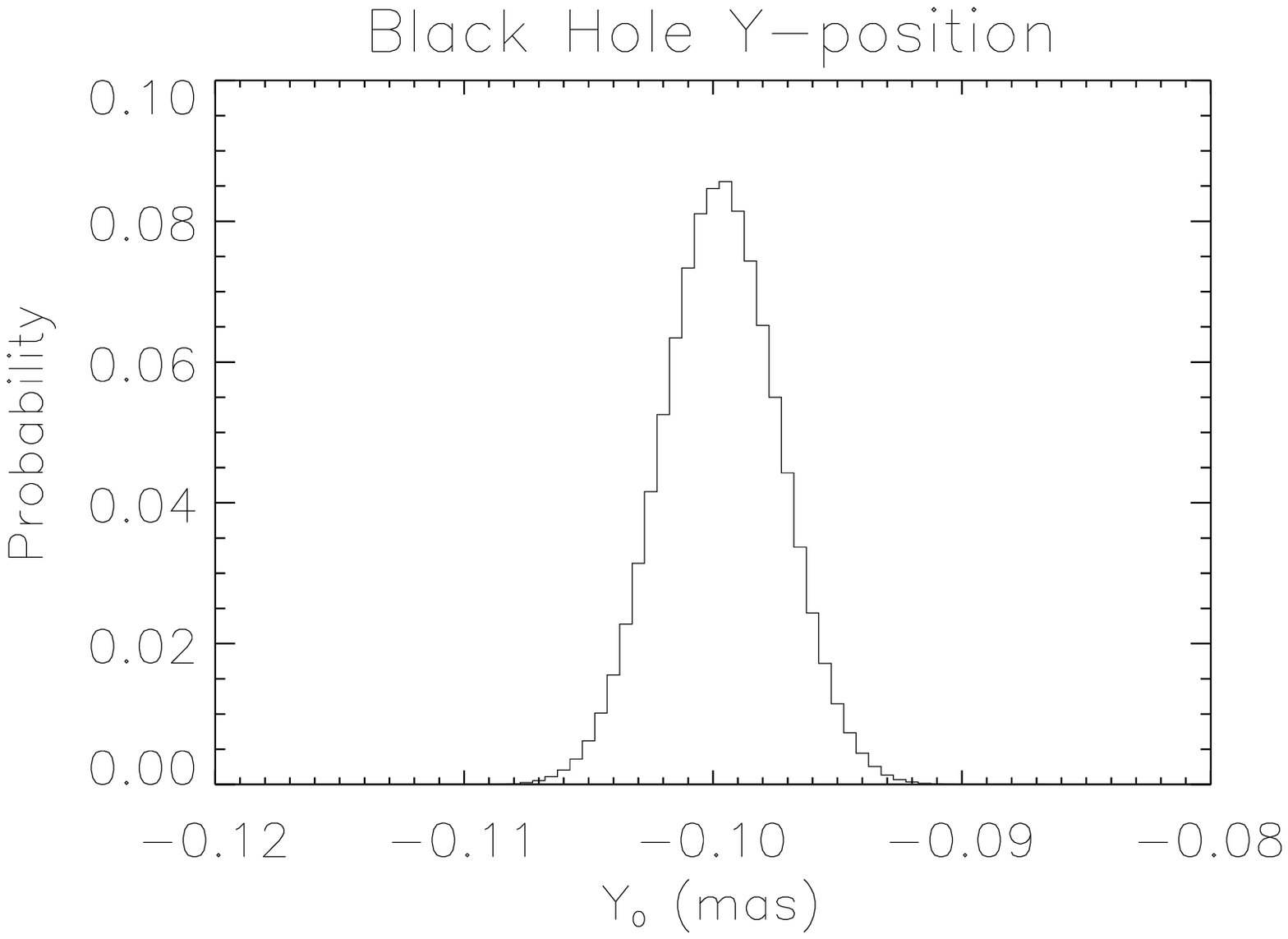}
\includegraphics[width=.32\textwidth]{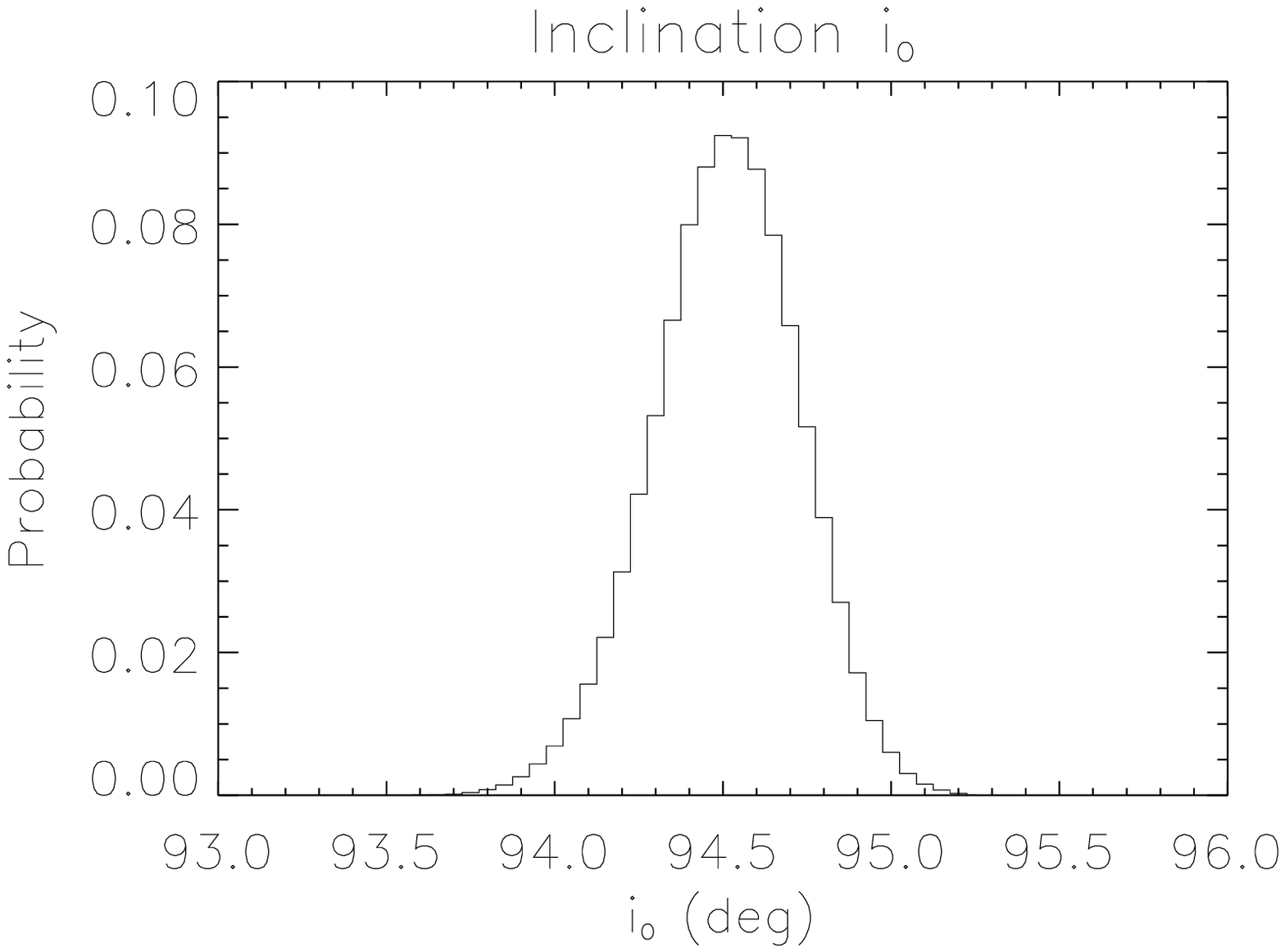}
\includegraphics[width=.32\textwidth]{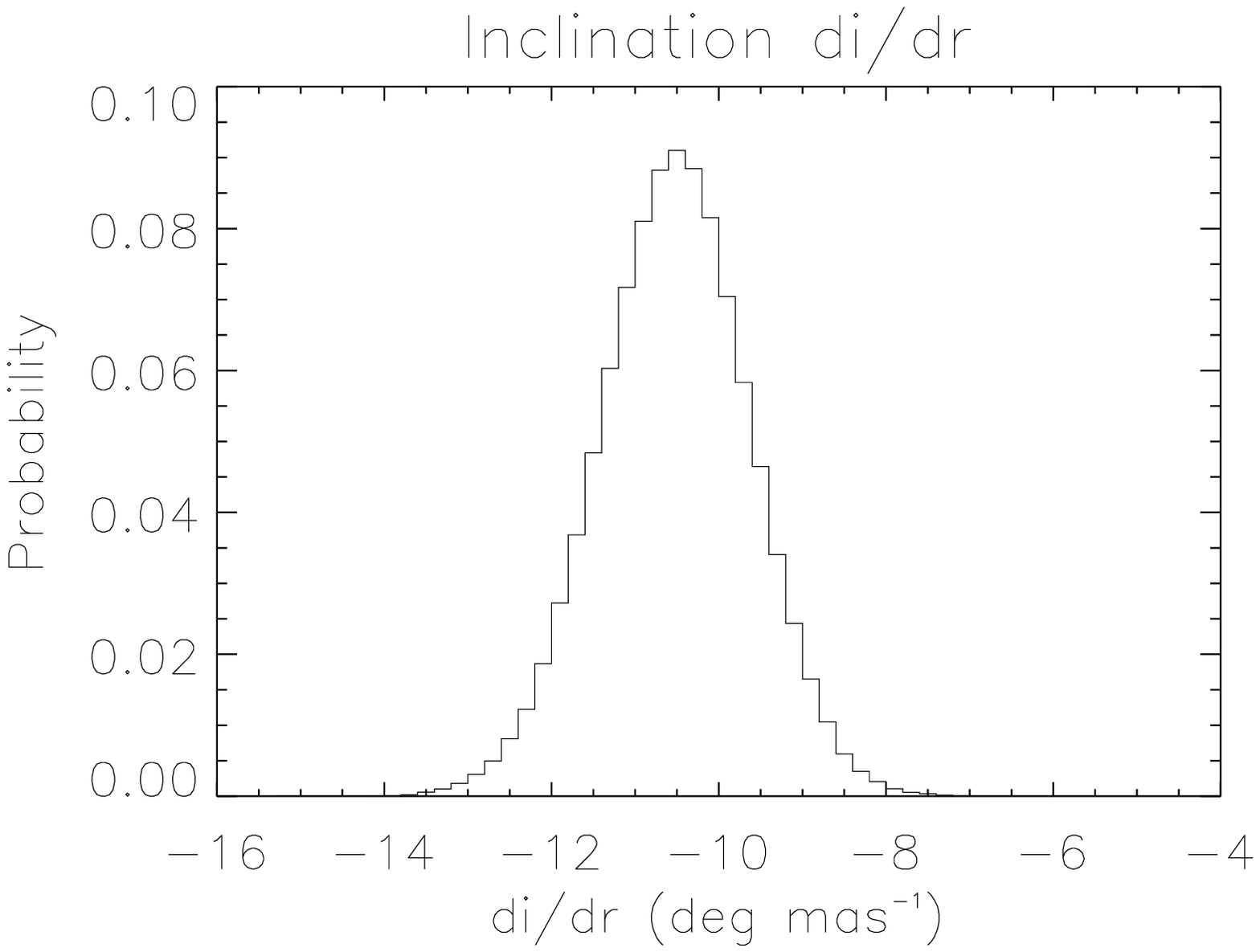}
\includegraphics[width=.32\textwidth]{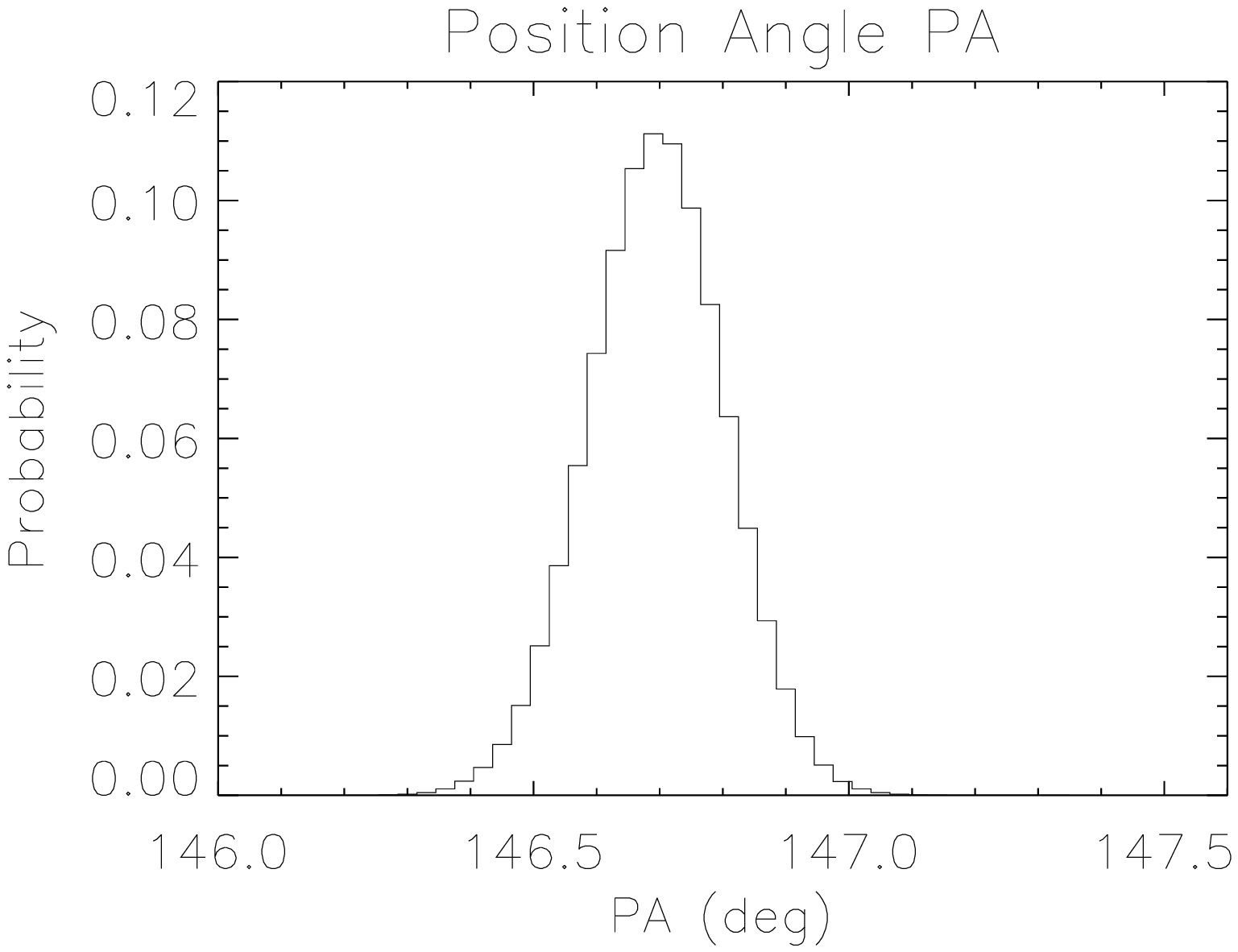}
\includegraphics[width=.32\textwidth]{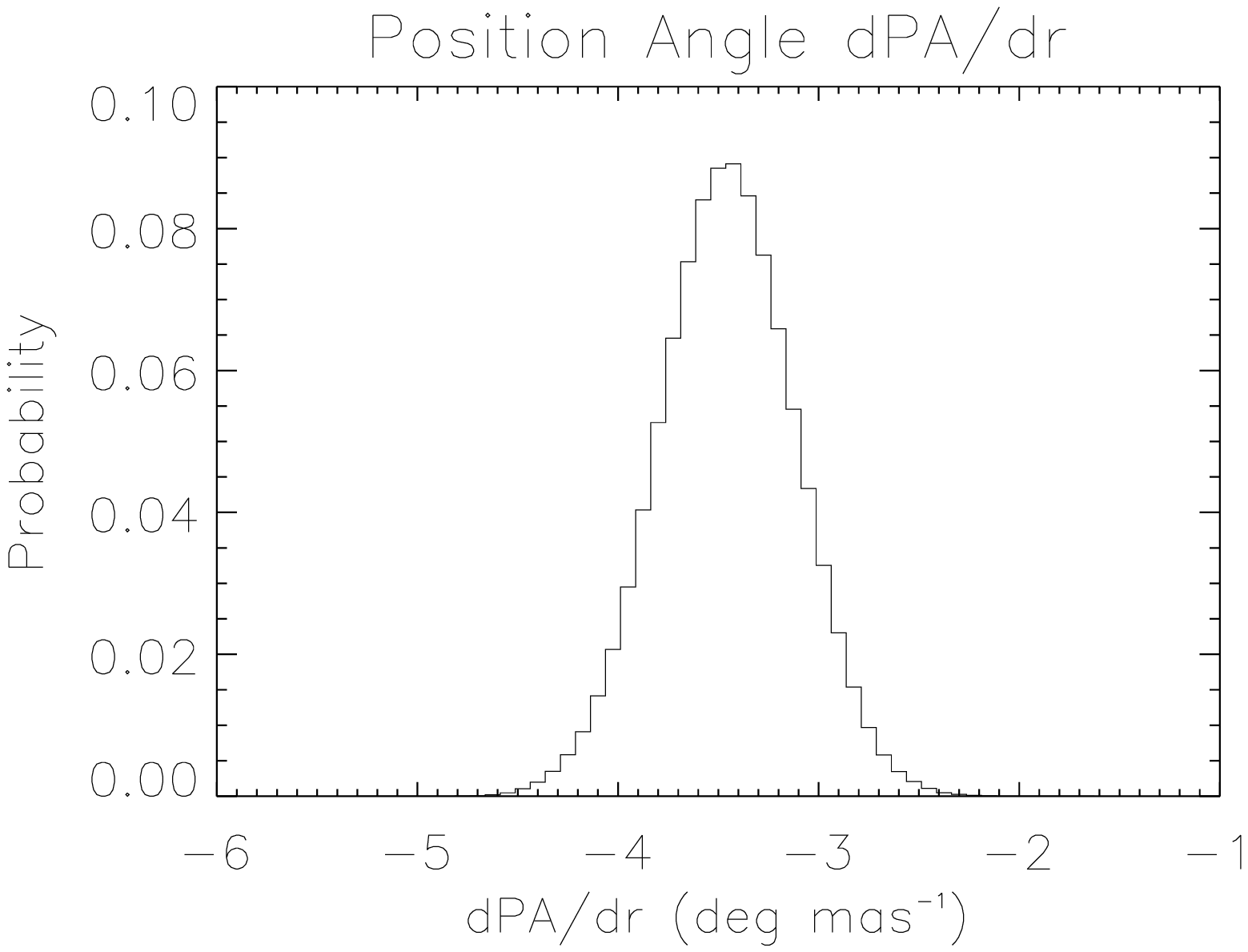}
\includegraphics[width=.32\textwidth]{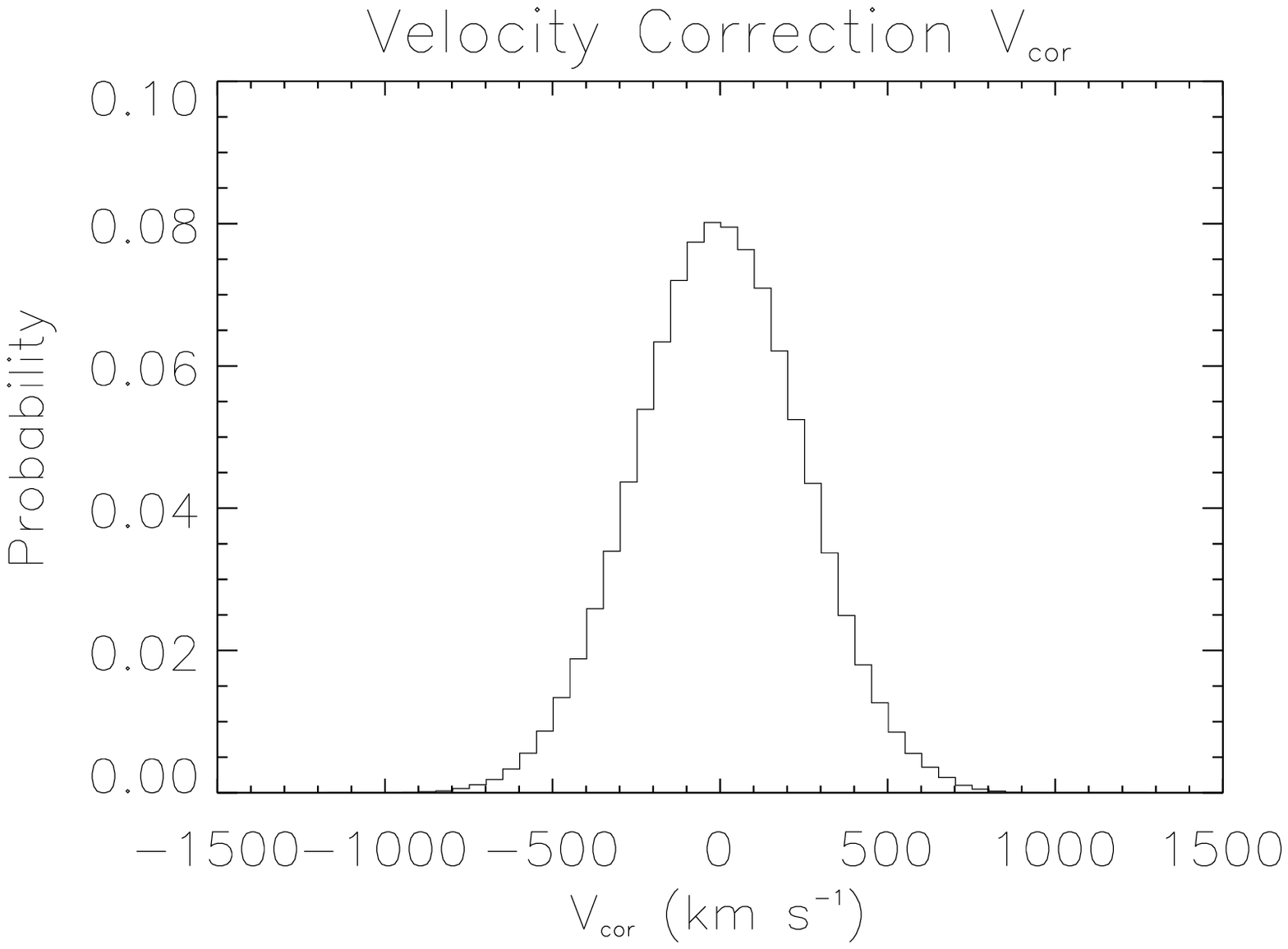}
\caption{\footnotesize 
Combined posteriori probability distributions for the global disk parameters describing the disk model for all 10 strands.
 \label{figure:HoPDF}
}
\end{figure}

\begin{figure}[h]
\epsscale{1.0}
\plotone{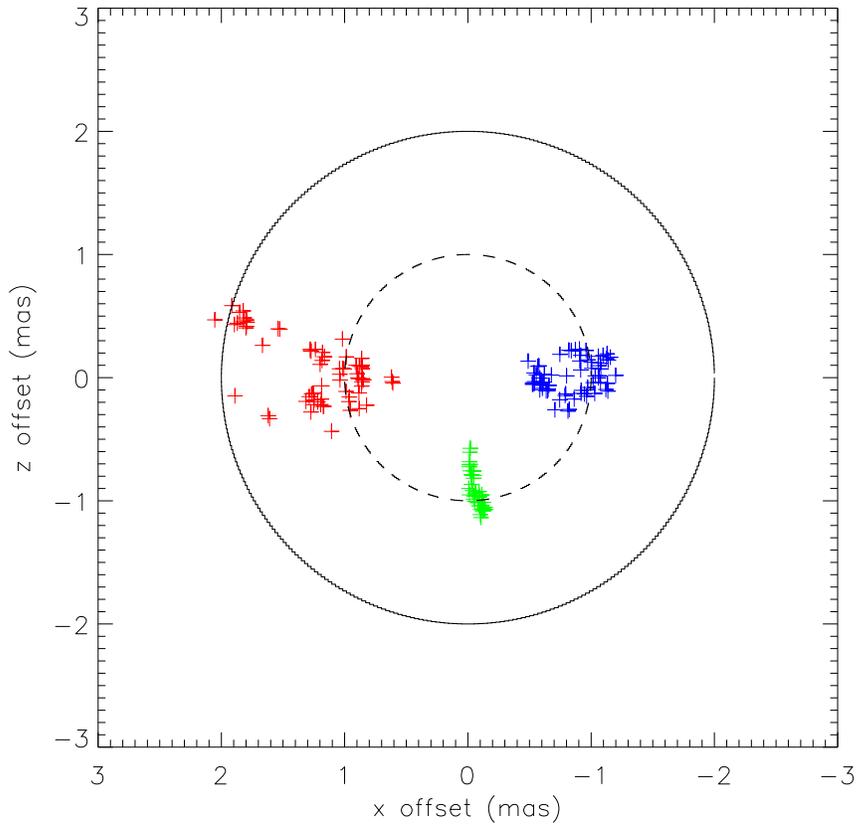} 
\caption{\footnotesize 
Model-fitted position distribution of the maser spots viewed on the
disk plane. The red, green and blue colors represent the red-shifted,
systemic and blue-shifted masers, respectively.  For the
high-velocity features (especially the red-shifted features), maser
spots are clustered into several clumps on the disk. This is different
from the case of NGC 4258 where the high-velocity masers are seen to be
radially localized (as shown in Humphreys et al. 2013).
 \label{figure:XZ}
}
\end{figure}

To come to our final result, we dropped the first 20\% of trials in each strand
to avoid using any un-converged trials in the final probability distribution.
For each of the global parameters,
we quote the mean value in the probability distribution function as
the fitted result, and 68\% confidence range as the error. All
uncertainties listed here have been scaled by $\sqrt{\chi^{2}/N}$,
i.e., $\sqrt{1.575}$. For $\Ho$, we determine 66.0 $\pm$ 5.0
  km~s$^{-1}$ Mpc$^{-1}$. We list the full set of final fitting results in
Table~\ref{table:Bayesian fitting result}. We also plot the
probability distribution for each global parameter in
Fig.~\ref{figure:HoPDF}, and show the distribution of maser features
on the disk in Fig.~\ref{figure:XZ}.

\subsection{Systematic uncertainty in $\Ho$ }
	
Apart from the formal fitting uncertainty for $\Ho$, here we
investigate several factors that might cause systematic uncertainty.
	
To test for human bias
in the acceleration measurement, here we fit the maser disk using accelerations
determined using method 2, rather than method 1, while still using the
difference between the two methods as the error floor for
acceleration data. This fit gives $\Ho = 68.5 \pm 6.0$ 
km~s$^{-1}$ Mpc$^{-1}$, which is consistent within
1-$\sigma$ of our original result.
	
For the systemic features, the error floor we used for the
acceleration data is the measured difference between results using
method 1 and method 2, for each maser spot independently. To test
the robustness of our measurement against the value of the error floor,
we also tried setting the error floor to
the mean difference between method 1 and method 2,
calculated separately for clump 1 and clump 2.
The mean difference in clump 1 is 0.045 km~s$^{-1}$
yr$^{-1}$, and for clump 2 the value is 0.20 km~s$^{-1}$ yr$^{-1}$. We
re-run the Bayesian fitting with other parameters unchanged and get
$\Ho = 68.0 \pm 6.4$ km~s$^{-1}$ Mpc$^{-1}$, 
consistent with our original result.
	
The error floor we used for the acceleration of high-velocity features
is the mean of the rms values measured for the red-shifted and
blue-shifted features. The rms for the red-shifted features is 0.155
km~s$^{-1}$ yr$^{-1}$, while the blue-shifted features had an rms of
0.275 km~s$^{-1}$ yr$^{-1}$. If we use 0.155 km~s$^{-1}$ yr$^{-1}$ as
the error floor only for the red-shifted masers and 0.275 km~s$^{-1}$
yr$^{-1}$ only for the blue-shifted masers, we get $\Ho = 66.5 \pm
5.7$ km~s$^{-1}$ Mpc$^{-1}$, which is consistent to within 
1-$\sigma$ of our original result of 66.0 $\pm$ 5.0
km~s$^{-1}$ Mpc$^{-1}$.
		
The clumpiness of the high-velocity masers on the VLBI map suggests the
existence of spiral structure in the accretion disk of NGC 5765b (see
the next section for a detailed discussion). This phenomenon has
been discussed for the maser disk of NGC 4258 (Humphreys et al. 2013),
in which the unmodeled spiral structure is claimed to contribute 1\%
uncertainty for the final distance determination. Here we simply
estimate that spiral structure in NGC 57675b similarly contributes 1\%
to the final error budget.
	
After including these sources of systematic uncertainty, our final
measurement gives $\Ho$ of 66.0 $\pm$ 6.0 km~s$^{-1}$ Mpc$^{-1}$. We
list all the detailed uncertainties in
Table~\ref{table:H0uncertainty}.

\section{DISCUSSION}

\subsection{Essential requirements for good distance measurements}
	
Combined with our previous experience, we note that a good distance
measurement (with uncertainty less than 10 \%) using the ``disk maser"
method requires both a well-defined rotation curve from the
high-velocity maser components and well-measured
accelerations over a wide velocity range for the systemic masers. In
principle, only one systemic feature with a perfectly measured
acceleration could provide the distance, since the rotation curve
defined by the high-velocity features measures the enclosed mass, the
dynamic center and recession velocity. In practice, a large spread of
systemic features over a wide velocity range not only reduces the mean
acceleration measurement uncertainty, but also traces the inclination
warping with masers at different radii. The case of NGC 5765b fulfills
such requirements in the following aspects:
	
1. The system has bright maser lines, benefitting VLBI calibration and imaging.
	
A minimum maser line flux of $\sim$ 100 mJy is required to get a solid
fringe detection for the VLBA+GBT array, within 1-2 mins integration
time and 125 kHz channel size under moderate atmospheric
conditions. For the VLBA+GBT+VLA array, this limit goes down to $\sim$
50 mJy. In the case of NGC 5765b, the strongest maser line is about
200 mJy, which is ideal to provide sufficient calibration of the
atmospheric and instrumental effects.
	
As mentioned in Section 2.2, the relative position uncertainty of each maser
spot is approximately given by 0.5$\Theta$/(S/N). For NGC 5765b, the
blue-shifted features are the faintest, with most of them being below
15 mJy (as shown in Fig.~\ref{figure:gbtspec}). The systemic masers
for which the acceleration has been measured are almost all above 20 mJy. The 1$\sigma$
noise level from our VLBI map is 0.3 mJy beam$^{-1}$ channel$^{-1}$ (with a
channel width of 125 kHz), providing a 10$\sigma$
measurement of features at 3 mJy. Compare this to the Kuo et
al. (2015) result on NGC 6323, where the systemic masers (which are
weaker than the high-velocity masers in that galaxy) are all
below 10 mJy. The authors conclude their final uncertainty ($\sim$ 30
\%) on the distance measurement is dominated by the low precision of the
maser position measurements, even though their VLBI map has a noise
level similar to that in our study.
	
It is also advantageous that our VLBI observations were scheduled
closely in time (within 2 months) in both 2013 and 2014, which limits
the impact of maser variability when combining tracks.
						
2. The maser covers a wide velocity range in both the systemic and 
high-velocity masers.
	
As mentioned in Section 2.2, the rotation curve traced by the
high-velocity features defines M/D, where M is the enclosed mass and D
is the distance, while the slope of the systemic features on the P-V
diagram constrains $M/D^{3}$. So a wider velocity coverage on both the
systemic and high-velocity features provides better leverage and hence
better constraints on M and D. In the case of NGC 5765b, the
red-shifted and blue-shifted masers span over 350 km~s$^{-1}$ and 250
km~s$^{-1}$ respectively, and sample the rotation curve very well,
as shown in Fig.~\ref{figure:pv-map}.

The systemic masers in NGC 5765b span 90 km~s$^{-1}$, and we are able to measure
the acceleration of each line with a typical uncertainty of $\sim$10\% within 70 km~s$^{-1}$ velocity range. 
For comparison, in NGC 4258 the systemic features span over 110
km~s$^{-1}$(Humphreys et al. 2008), contributing to the
3\% distance measurement (Humphreys et al. 2013).
		
3. Accelerations of the systemic features are measured accurately.
	
In NGC 5765b, the low variability among the systemic features and
relatively low ($<$ 1.5 km~s$^{-1}$ yr$^{-1}$) but similar drift
rate for the majority of features allows us to trace them through all
observing epochs and contributes to our precise acceleration
measurements. However, the monitoring must span a longer time to 
measure these low accelerations.
The lifetimes of individual maser components in other galaxies varies
from a few
months to a few years. In NGC 5765b the systemic masers have shown
little structural changes over the observed 2.5 years, which has
allowed us to utilize all 2.5 years of monitoring data to analyze each
component.

We also note that the low variability and small accelerations in NGC 5765b
reduce the impact of the summertime gap with GBT monitoring data.
For maser galaxies with high accelerations or variability, it would be
better to fill the summer gap for the best acceleration measurements.
        
\subsection{Improving the $\Ho$ estimate with future NGC 5765b observations}
	
In this section we investigate chances to improve the $\Ho$
measurement with observations of NGC 5765b in the future. The constraint on $\Ho$ from
this work is 9.1 \%, with random uncertainties dominating the
final result rather than systematic uncertainties, as listed in
Table~\ref{table:H0uncertainty}. So in principle, the final result
could be improved by more observations. Since we use the same
methodology for our distance measurement as Humphreys et al. (2013)
for NGC 4258, which reached a 2.2 \% formal fitting uncertainty and a
3 \% total uncertainty, we compare the input data for both NGC 4258
and NGC 5765b quantitatively in this section.
	
The input data for the Bayesian fitting program we used here are the
sky position, velocity and acceleration with individual uncertainties
for each maser spot. The input data can be improved in two ways, by
observing more maser spots covering a wider velocity range or by
smaller uncertainties for each type of data.
	
We list the details of input data for both NGC 4258 and NGC 5765b in
Table~\ref{table:datacomparison} for comparison. Regarding the
velocity coverage, the most noticeable difference is the velocity
range for the systemic part with accelerations measured: 110
km~s$^{-1}$ in NGC 4258 versus 70 km~s$^{-1}$ in NGC 5765b. For the
high-velocity features, the velocity coverage is similar, which means
a similar sampling range of the Keplerian rotation curve as NGC
4258 and NGC 5765b have similar black hole masses. Due to the relatively
high intensity of the systemic masers in NGC 5765b and the noise level
we reached here (1$\sigma$ of 0.3 mJy beam$^{-1}$ channel$^{-1}$ (125 kHz)
for the VLBI map, and $<$ 2 mJy channel$^{-1}$ (24.4 kHz) for the single dish monitoring
spectra), increasing either VLBI or single dish observing
sensitivity will not substantially cover new velocities.
So as a practical matter, it would be difficult to
improve the overall fitting result by including more maser spots.
	
On the other hand, regarding the uncertainty for each type of data,
the most noticeable difference between NGC 5765b and NGC 4258 lies in
the acceleration data. For NGC 4258, the uncertainty on acceleration
for each measured maser component is less than 4\% (Humphreys et al. 2008),
while for NGC 5765b the typical uncertainty is 10\% for each maser component. The VLBI
position uncertainties are similar between NGC 4258 and NGC 5765b,
after adding the position error floor.  So we conclude that future
improvement of the $\Ho$ uncertainty with NGC 5765b depends on measuring accelerations
with better precision.

\subsection{Clumpiness of the high-velocity masers}
A visual inspection of the GBT spectrum (insets in
Fig.~\ref{figure:gbtspec}) of the high-velocity components reveals the
maser lines are regularly clustered into several clumps. This is
especially obvious on the red-shifted features, in which there are
mainly five clumps evenly distributed between 8740 km~s$^{-1}$ and
8980 km~s$^{-1}$, spaced by $\sim$50
km~s$^{-1}$. Beyond this, there is another clump centered at 9075
km~s$^{-1}$.
	
\begin{figure}[h]
\epsscale{1.0}
\plotone{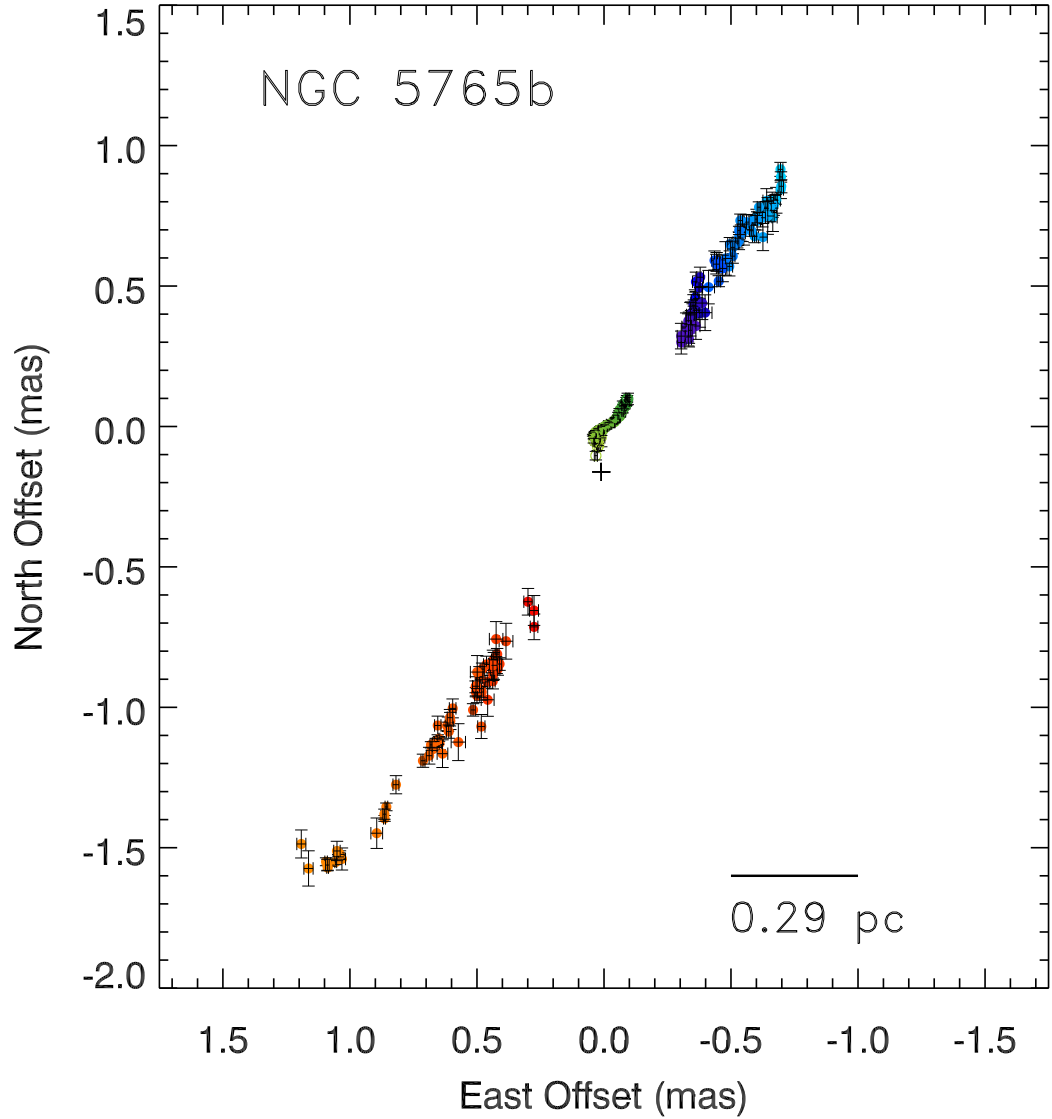} 
\caption{\footnotesize 
The combined VLBI map with a flux cutoff of 3 mJy, which is at the
10$\sigma$ level. The maser positions are referenced to the
strongest systemic feature at 8295.91 km~s$^{-1}$, with the black
cross marking the dynamic center derived from the fitting of the
rotation curve, as shown in Fig.~\ref{figure:pv-map}. The horizontal
bar in the lower right corner marks the corresponding physical size,
using a nominal distance of 119 Mpc to NGC 5765b.
\label{figure:vlbi-map}
}
\end{figure}

The VLBI map (Fig.~\ref{figure:vlbi-map}) shows that the red-shifted masers
are also spatially clustered into several clumps, with a typical
spacing between clumps of $\sim$0.3 mas (0.18 pc).  The 
blue-shifted masers, meanwhile, are distributed more smoothly and mainly
appear in 2 clumps.
		
Such clumpiness of the high-velocity masers has been seen in several
disk maser systems (e.g. NGC 4258 by Humphreys et al. (2008) (2013);
UGC 3789 and NGC 2960 by Kuo et al. (2011)) but only discussed in
detail for NGC 4258. We note that the spacing of clumps appears to 
be similar between the red and blue side of the disk only in the case
of NGC 4258.  Another thing to note is that the features in NGC 4258 appear clumped in
radial arcs as shown in Humphreys et al. (2013), while for NGC 5765b, they appear clumped in both radius and azimuthal angle, as shown in
Fig.~\ref{figure:XZ}.

Several models have been proposed to explain the generation of such
clumpiness, including a spiral density wave model (Maoz 1995) and
a spiral shock model (Maoz \& McKee 1998).
	
The spiral shock model proposed that masers are generated in
spiral shock regions and this could explain the asymmetry of the
intensity between the red-shifted and blue-shifted features, as first
seen in NGC 4258. This model also predicts that the red-shifted
features will have negative accelerations, while the blue-shifted
features have positive accelerations. In the case of NGC 5765b, the
weighted average accelerations of the high-velocity features are
(red-shifted) 0.008 $\pm$ 0.155 km~s$^{-1}$~yr$^{-1}$ and
(blue-shifted) -0.049 $\pm$ 0.275 km~s$^{-1}$~yr$^{-1}$. These results
are consistent with zero acceleration, though with large
measurement uncertainties. One particular blue-shifted
feature at 7582.95 km~s$^{-1}$ has an acceleration of -0.528 $\pm$
0.110 km~s$^{-1}$~yr$^{-1}$, which is inconsistent with predictions
from the spiral shock model. So even though the red-shifted features
are brighter than the blue-shifted features in NGC 5765b, our
acceleration results do not support the spiral shock model. In a
parallel MCP paper, Pesce et al. (2015) systematically examine the
predictions from this model using all the disk maser spectra
obtained from the MCP, and they do not find clear evidence supporting
this model.
	
The spiral density wave model proposed that spiral structures are
generated from nonaxisymmetric perturbations on the accretion disk and
maser clumps are located at the intersection of the spiral arms and
the disk mid-line. So the high-velocity masers would be distributed
within a few degrees from the mid-line and would have very small LOS
accelerations that can be slightly positive or negative. This model
works well for NGC 4258, where the maser clumps for both the red-shifted and
blue-shifted features are periodic in disk
radius with the same characteristic scale of $\sim$ 0.75 mas (\eg
Humphreys et al. 2008). In the case of NGC 5765b, even though
the red-shifted features are clustered into several clumps with an
spacing of $\sim$0.3 mas, the blue-shifted features do not
show the same spacing (as shown in Fig.~\ref{figure:vlbi-map}). So
more precise measurements are needed to test the spiral density
wave model in NGC 5765b.
	
\subsection{Amplitude vs. position for the high-velocity masers}
	
Bragg et al. (2000) compared the maser flux with maser position in NGC
4258, where no dependence between line amplitude and radius was
found. Instead, they found that the maser amplitude peaks around the midline. This
was interpreted as masers originate in aligned clumps of material that
amplify each other, so the observed amplitude is largely independent
of radius.
	
\begin{figure}[h]
\epsscale{1.0}
\includegraphics[width=.5\textwidth]{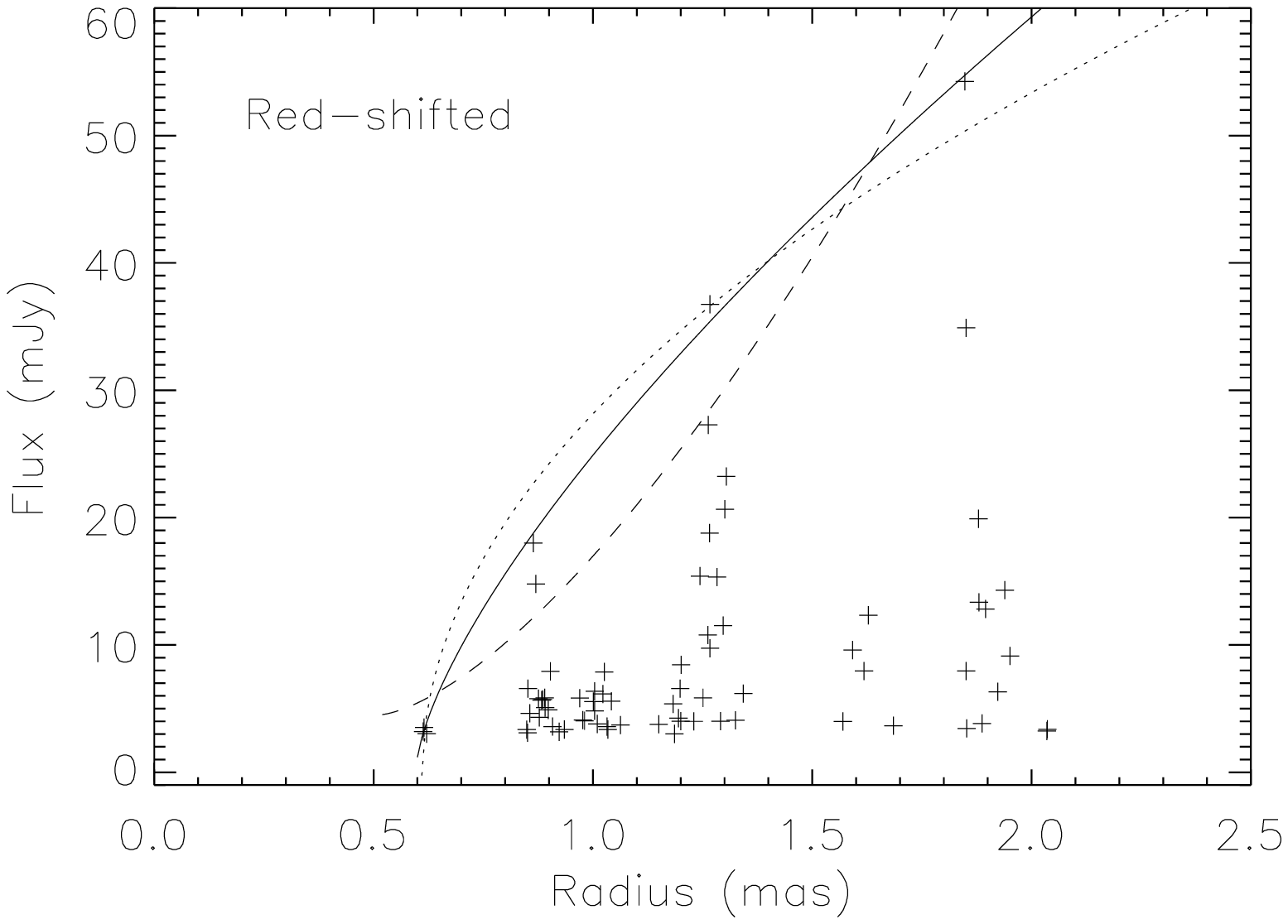} 
\includegraphics[width=.5\textwidth]{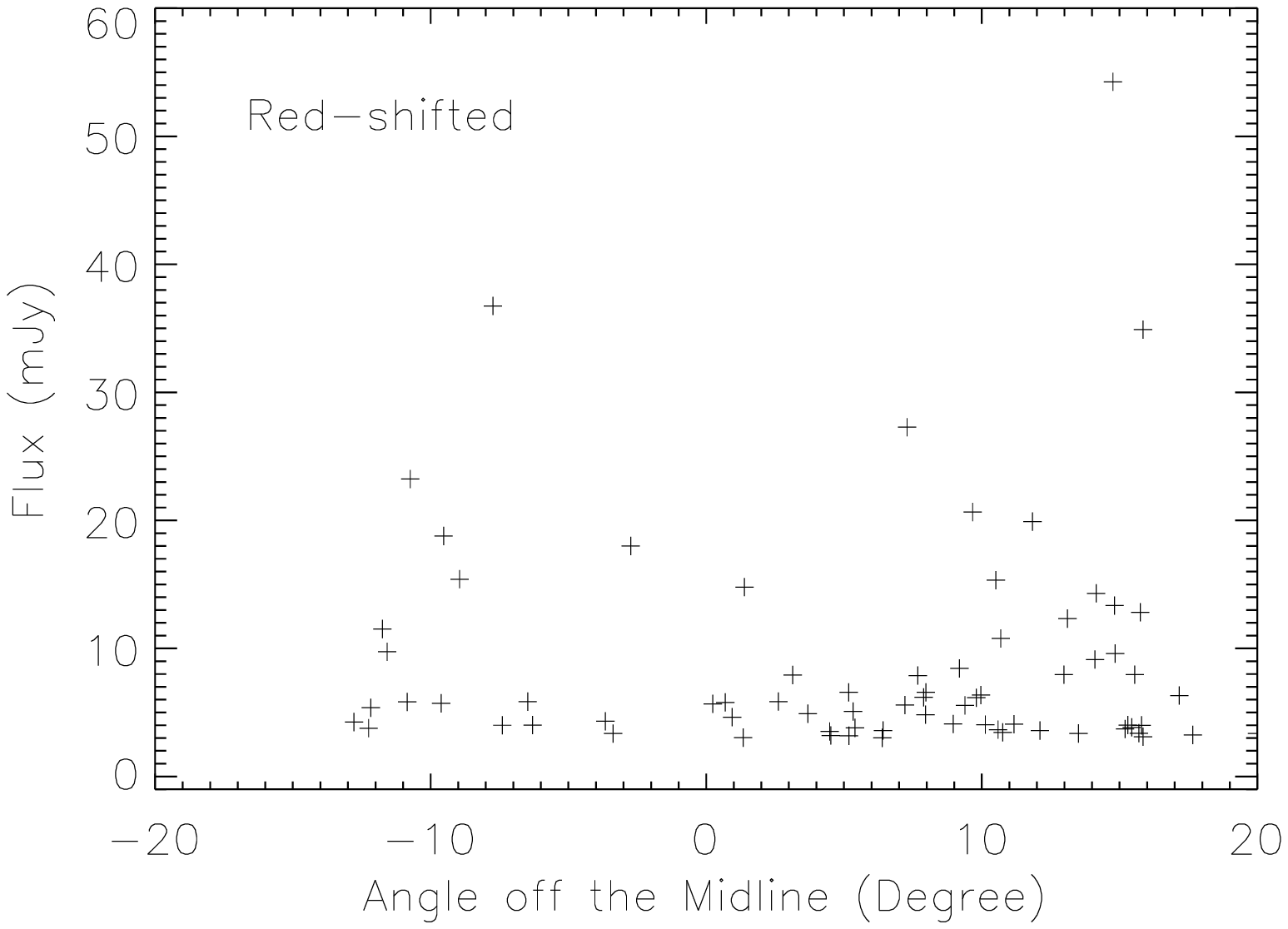} 
\includegraphics[width=.5\textwidth]{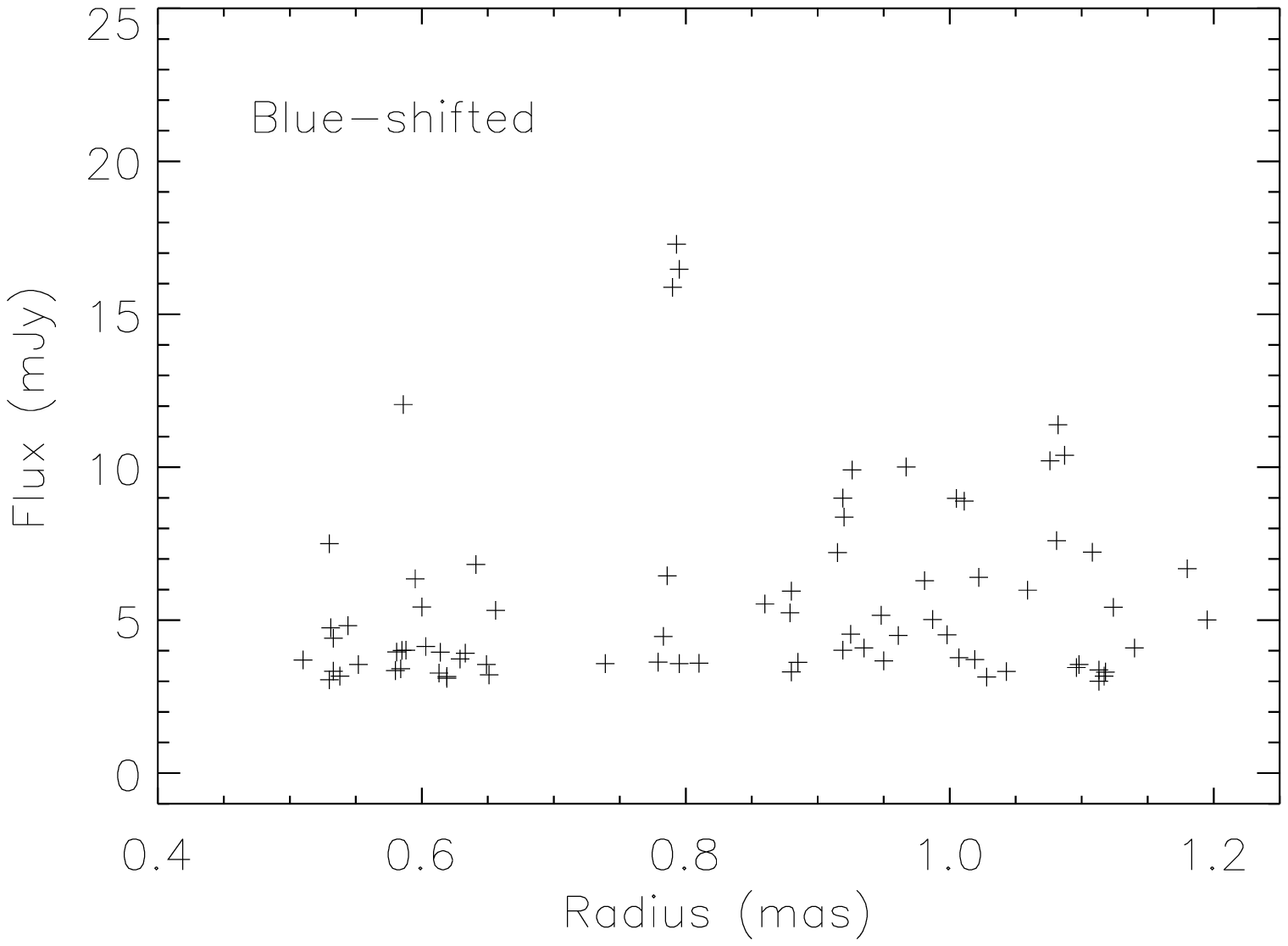} 
\includegraphics[width=.5\textwidth]{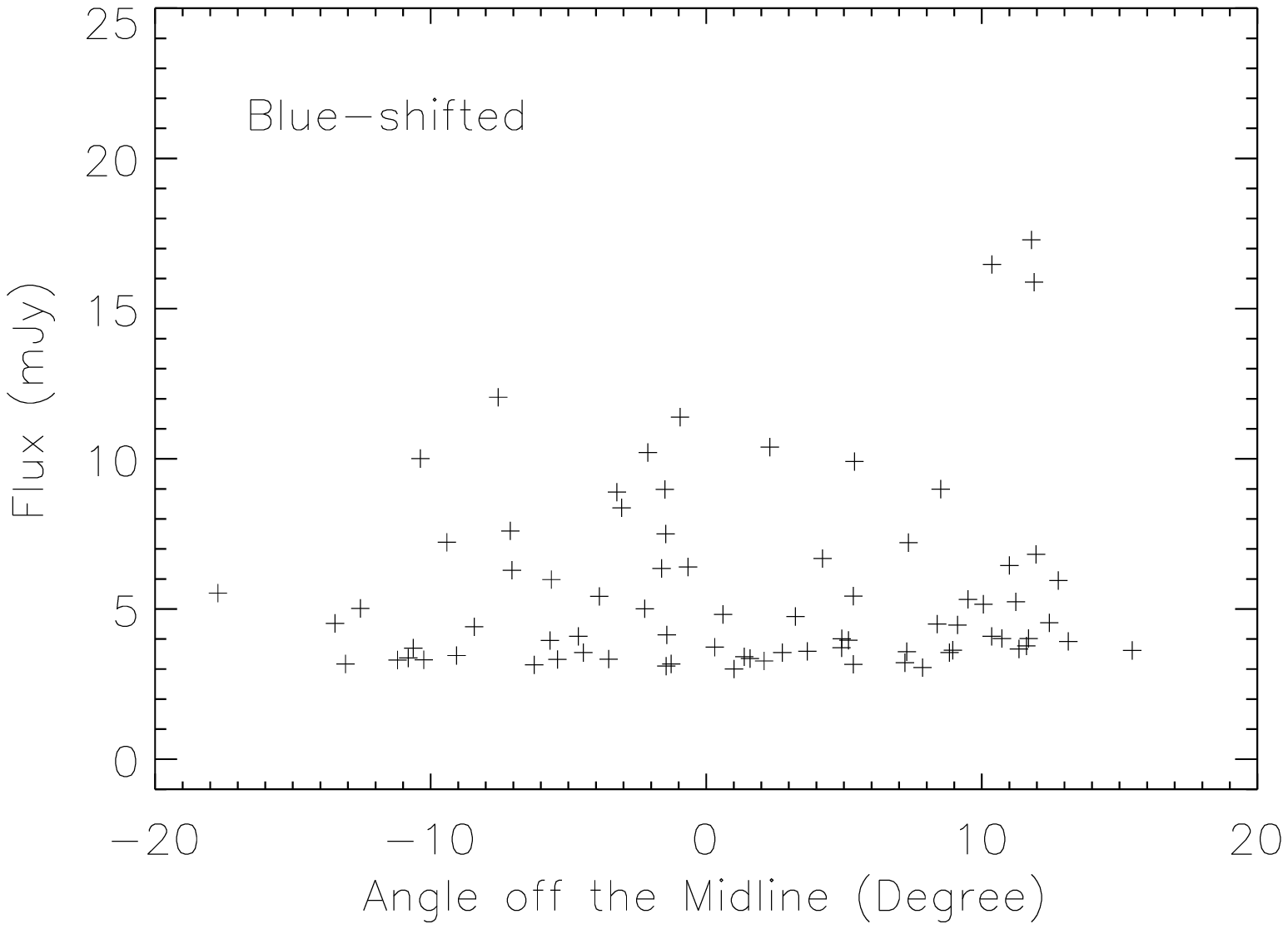} 
\caption{\footnotesize 
Modeled radius and angle versus peak intensity for the high-velocity
masers in NGC 5765b. The upper left plot shows the radius versus peak
intensity for the red-shifted features, and the solid line shows the best
fit to the outline of the maser line peaks, with flux $\propto$
r$^{0.7}$. For comparison, the dotted line has flux $\propto$
r$^{0.5}$, while the dashed line has flux $\propto$ r$^{1.5}$. The
upper right panel shows the angle off the midline versus maser peak
intensity for the red-shifted features. The maser peaks appear 
away from the midline. Plots for the blue-shifted
features are on the bottom panel. The blue-shifted features
show no such dependencies as seen in the red-shifted features.
 \label{figure:rflux}
}
\end{figure}

To test these results for NGC 5765b, we plot the maser amplitude
versus disk radius and versus deviation from the mid-line for the
high-velocity masers in NGC 5765b in Fig.~\ref{figure:rflux}. The flux
comes from the VLBI map, which represents the mean maser
intensity over our VLBI observations. The radius comes from the
Bayesian fitting result for each maser spot.
	
It is interesting that for the red-shifted lines, the peak intensities
of maser lines are proportional to their radius, and masers peak
away from the midline. We use a power-law function to fit the outline
of the peak intensities versus radius for the red-shifted features and
the best fit yields a power index of 0.70 $\pm$ 0.050
(1$\sigma$), as shown by the solid line on the plot.
	
To see whether our fitted results are specific to the VLBI data only,
we also use the GBT monitoring data for a consistency check: We
measure the peak intensity closest (within 1 km~s$^{-1}$) to the
velocities where the VLBI intensity peaks, namely 8753.8
km~s$^{-1}$, 8856.0 km~s$^{-1}$, 8968.5 km~s$^{-1}$, 9077.5
km~s$^{-1}$, 9079.0 km~s$^{-1}$ and 9081.3 km~s$^{-1}$, and use these
data to re-fit the power index for each epoch of the monitoring
observations. The mean power law index is 0.75 $\pm$ 0.014
(1$\sigma$). So the power index we measured from the VLBI data is
consistent with this result to within 3$\sigma$.
	
Intuitively, this positive flux-radius dependency seems natural.
The LOS velocity of the high-velocity masers goes as $V_{rot} \cdot
cos(\phi)$, where $V_{rot}$ is the Keplerian rotation speed and $\phi$
is the angular deviation from the mid-line. To maintain a velocity
coherence of $\sim$ 1 km~s$^{-1}$ required for maser amplification,
maser spots with higher rotation velocities (thus closer to the
dynamic center) would have shorter gain length. This would result in
masers at higher rotation velocity having lower intensity by assuming
the maser intensities are proportional to the gain length. If we
assume the maser is saturated, a detailed calculation (see 
Appendix~\ref{b:amp-radius}) shows the maser flux goes with radius as
f $\propto$ r$^{1.5}$ with the maser flux linearly proportional to the
gain length. For comparison, we plot the fitted curve with the power
index fixed to 0.5 (dotted line) and 1.5 (dashed line) in
Fig.~\ref{figure:rflux} (a).
	
Since the fitted power index is inconsistent with the calculation
above, a better explanation is needed. The maser flux may not scale
with the gain length linearly, as the maser might be partially saturated.
 Also we didn't include the disk warping
in our calculation, which would re-shape the regions where maser flux
peaks. The disk warping may also cause the line peaks to deviate
from the mid-line. Another factor that should be taken into account is
the different gas and dust temperatures at different radii, which
would have a significant effect on the maser pumping efficiency (\eg
Yates, et al. (1997)).
		 		
Pesce et al. (2015) compiled a list of 32
clean\footnote{Here ``clean" disk means that maser emission
generated from the Keplerian disk dominates over that generated
from jets or outflows, and the disk itself displays Keplerian rotation.}
 megamaser disk systems discovered so far
and present their time-averaged single dish spectra. 
Six of these maser systems show
a similar dependency of maser flux on radius as we have seen
in the red-shifted part of NGC 5765b here. They are: J0437+2456
(blue-shifted), UGC 3789 (blue-shifted), ESO269-G012 (blue-shifted),
UGC 9639 (blue-shifted), NGC 6264 (red-shifted) and CGCG498-038
(red-shifted). 
	 
\subsection{Beaming angle of the systemic masers}
The slope of the systemic features on the P-V diagram is related to
the physical radius of those spots to the dynamic center ($
\sqrt{GM/r^{3}}$). From the P-V diagram, we fit a slope of 435.7
km~s$^{-1}$ mas$^{-1}$. Combined with the black hole mass of $4.4
\times10^{7}M_\odot˜$ and nominal distance of 119 Mpc, this result
corresponds to 0.68 pc. The Keplerian rotation velocity at this radius
is about 530 km~s$^{-1}$.

The systemic masers span 90 km~s$^{-1}$ in NGC 5765b (from 8240 to
8330 km~s$^{-1}$). For simplicity, if we assume all the systemic
features are located at the same radius, this would give a beaming
angle of 9.8$^{\circ}$ on the azimuth direction for the systemic
masers. However, since not all the systemic masers lie on the same
radius, as shown by our Bayesian fitting result, here instead we use
the velocity range of 8260 to 8290 km~s$^{-1}$, in which masers are
located at a similar radius of about 1 mas on the disk, to estimate
the lower limit of the beaming angle. This gives an azimuthal beaming angle of
3.3$^{\circ}$.

For comparison, among the other published megamaser disk systems,
only two have azimuthal beaming angles estimated.  NGC 4258 has a beaming
angle of 7$^{\circ}$, measured both by the slope of the
systemic features (Miyoshi et al. 1995) and by monitoring the
drift of systemic features's velocity with time (Moran 2008). For
UGC 3789, based on the slope of the systemic features on the P-V
diagram by Braatz et al.(2010), we estimate the
beaming angle to be 4.4$^{\circ}$ and 2.5$^{\circ}$ respectively for
the two maser rings, as described in Braatz et al. (2010).
		
\subsection{ The profile and variability of systemic masers in NGC 5765b compared to NGC 4258}	
The profile of the systemic masers in NGC 5765b is smooth compared to
the ``spiky" high-velocity masers. This contrast is more obvious than
the case of NGC 4258. One might consider whether the blended/``spiky"
difference arises because the systemic features are amplifying the
background radio continuum emission. Also since the systemic features
all have similar LOS velocity, their velocities overlap and the total
spectral profile is likely the sum of weak maser emission coming from
numerous gas clouds in the accretion disk. The high-velocity masers
meanwhile are self-amplified, and the spikiness may reflect a sparse
availability of seed photons and different gain paths.
		
Another feature of the systemic maser spectrum is that there is
little change of the overall profile across the observed 2.5 yrs as
shown in Fig.~\ref{figure:stack-sys}, unlike the case of NGC
4258, which has shown clear structural changes on a few-months time
scale (e.g. Argon et al. 2007). The ``notches'' and ``dents'' formed
between strong features in the systemic part in NGC 5765b (especially
between 8260 and 8290 km~s$^{-1}$) co-drift with the maser peaks.
This means the masing gas clouds are
distributed closely with each other spatially, and they lie on a
similar radius, which will have similar accelerations. This is
consistent with the Bayesian fitting results, in which masers in the
velocity range of 8260 and 8290 km~s$^{-1}$ are all located at radii
between 0.895 to 1.126 mas on the disk.
	
The black hole masses for NGC 4258 (4.0 $\times~10^{7}M_\odot˜$ from
Humphreys et al. 2013) and NGC 5765b (4.55 $\times~10^{7}M_\odot˜$
from this work) are similar. The systemic maser spots in NGC 4258 are
distributed around 0.14 pc in radius on the disk, while for NGC 5765b,
systemic masers cover a larger range of radii, from
0.39 to 0.64 pc. 
As mentioned by Wallin et al. (1999), the
change of the velocity field on the LOS direction may be the main
reason for the spectral variability, so maser spots at larger radii
would have a smaller change in the LOS velocity field in a certain
amount of time according to the Keplerian rotation, thus showing less
spectral variation. Qualitatively, the variabilities seen in NGC 4258
and NGC 5765b are consistent with this explanation.
	
Other explanations for maser variability include the change of the
local physical conditions for pumping the masers and variability of the
background continuum source which is particularly important if the masers 
are unsaturated. The
persistence of the systemic masers in NGC 5765b might therefore be a result of
a less variable background source, or an indication that the maser is saturated.
	
If the change of the LOS velocity field is the main reason for maser
variability, future observations might be able to detect a correlation
between maser variability and acceleration for the systemic features
in all disk maser systems, since the acceleration depends on
the radius.

\subsection{Searching for the background radio continuum emission}
As mentioned above, systemic masers in NGC 5765b may amplify the
background radio continuum emission. We searched for radio continuum
emission from the line-free channels in our VLBI dataset, but did not
detect continuum emission at an upper limit of 0.1 mJy at 22
GHz. However, our recent VLA observation of NGC 5765b reveals a
continuum flux of 5 mJy at 22 GHz. Also, a point source is detected with an
integrated flux of 13.45 mJy at 1.4 GHz from the FIRST survey. So
either the 22 GHz radio continuum emission associated with NGC 5765b
is compact and variable, falling below our detection limit during the
VLBI observations, or the emission has been resolved by the VLBI
observations.

\section{CONCLUSIONS}
We present here the geometric distance measurement to the megamaser
disk host galaxy NGC 5765b, as part of the Megamaser Cosmology Project
(MCP). Owing to the relatively strong maser emission ($>$ 200 mJy), new
VLBI capabilities, and slow but discrete accelerations of the systemic
masers, we are able to measure the distance with very high precision,
from which we determine $\Ho$ = 66.0 $\pm$ 6.0 km~s$^{-1}$ Mpc$^{-1}$ and an
enclosed mass of 4.55 $\pm$ 0.40 $\times~10^{7}M_\odot˜$. The case
of NGC 5765b demonstrates essential requirements for good distance
measurements ($<$ 10 \% uncertainty), which can serve as
guidelines for future distance measurement work.

We also investigate some of the physical properties of the maser disk in
NGC 5765b: the red-shifted and blue-shifted features are spatially clustered
into several clumps with different characteristic scales on the red and
blue sides. Combined with the acceleration measurements, these results
favor the spiral density wave model, while the spiral shock model
is not supported. We find that the peak maser amplitude depends on
radius in the red-shifted features.  NGC 5765b shows little variation of the
systemic maser profile over the two years of observation. This is
quite different compared to the case of NGC 4258. If variability
is dominated by the change of the LOS velocity field due to Keplerian
rotation, we expect to see a positive correlation between systemic
maser variability and acceleration. This hypothesis could be tested
within current disk maser systems.

\appendix

\section{Requirement on absolute source position accuracy in MCP VLBI observations} 
\label{a:positionaccuracy}
	
For a given baseline $B$, the interferometric phase of a target source
at direction $S$ and observing frequency $\nu_{o}$ is given by
$\phi_{o} = 2\pi \nu_{o} \frac{B \cdot S}{c}$. During data
cross-correlation, the estimated source position is taken into
account to remove the source geometric phase. After this step, the
phase becomes $\phi_{o} = 2\pi \nu_{o} \frac{B \cdot
  (S_{o}-S_{c})}{c}$, where ${S_{o}}$ is the true source position, and
${S_{c}}$ is the estimated source position used in data
cross-correlation. For VLBI observations of masers, the phase of a
target maser spot at frequency $\nu=\nu_{o}+\Delta\nu$ is:
\begin{eqnarray*}
\phi & = & 2\pi \nu \frac{B \cdot (S_{o}-S_{c}+S)}{c} \\
& = & 2\pi (\nu_{o}+\Delta\nu) \frac{B \cdot (S_{o}-S_{c}+S)}{c}
\end{eqnarray*}
\\ where $S$ is the projected position difference between the maser spot at frequency $\nu$ and the reference position ${S_{o}}$.
	
Subtracting the phase corresponding to the reference position from the
phase of a given maser spot, we got the relative phase difference
$\Delta\phi$, which is what we actually measured. $\Delta\phi$ is
given by:
	\begin{eqnarray*}
	\Delta\phi & = & \phi-\phi_{o} \\
	& = & 2\pi (\nu_{o}+\Delta\nu) \frac{B \cdot (S_{o}-S_{c}+S)}{c} -  2\pi \nu_{o} \frac{B \cdot S}{c} \\
	& = & 2\pi\frac{B}{c}  \lbrack (\nu_{o}+\Delta\nu)(S_{o}-S_{c}+S) - \nu_{o}(S_{o}-S_{c}) \rbrack \\
	& = & 2\pi\frac{B}{c}  \lbrack \nu_{o} S+\Delta\nu (S_{o}-S_{c}+S) \rbrack
	\end{eqnarray*}
	
The second term in the bracket contains the additional phase caused by
absolute source position uncertainty $\Delta\nu (S_{o}-S_{c})$. To
neglect this term's effect would require $\frac{\nu_{o} S}{\Delta\nu
  (S_{o}-S_{c}+S)} \gg 1$.
	
For NGC 5765b, the frequency for the reference feature is $\nu_{o}$ =
21636 MHz, and all high-velocity features are within 64 MHz of
the reference feature. Based on our initial VLBI map, the
position difference of the high-velocity features and the reference
feature in the systemic part is greater than 0.3 mas. Putting these
numbers in the equation above, if we require the additional phase to be
less than 1/10 of the original phase, we have
$(S_{o}-S_{c}+S)\sim(S_{o}-S_{c}) \le 10$ mas, so the absolute source
position uncertainty should be within 10 mas.
	
For other maser targets, the required position accuracy depends on the
maser frequency coverage and the size of the maser disk. For targets
in the MCP, 10 mas is usually sufficient.

\section{Derivation of the relation between maser intensity and radius on the disk}
\label{b:amp-radius}

Suppose maser clouds are located on the disk with position (r ,
$\phi$), in which $\phi$ is measured relative to the LOS direction,
and on the midline red-shifted masers have $\phi = +$90, while
blue-shifted masers have $\phi = -$90. For an arbitrary maser cloud,
the LOS velocity is given by V$_{1}$ = V$_{r} \cdot cos(90-\phi_{1})$, 
where V$_{r}$ is the Keplerian rotation velocity
defined by V$_{r} = \sqrt {GM/r}$. For another masing cloud located at
(r , $\phi_{2}$ = $\phi_{1}+\Delta\phi$), the LOS velocity is given by
V$_{2}$ = V$_{r} \cdot cos(90-\phi_{1}-\Delta\phi) $. Maintaining the
velocity coherence required for maser amplification would require
$\Delta$V = V$_{2} -$ V$_{1} \le$ 1 km~s$^{-1}$. If $\Delta\phi$ is
infinitesimal, $\Delta$V can be written as $\Delta$V = $\sqrt {GM/r}
\cdot cos(\phi_{1}) \cdot \Delta\phi \le$ 1 km~s$^{-1}$.
	
For the same maser clouds, we define the gain length as the position
difference projected on the LOS direction $\Delta$L = r $ \cdot
(cos\phi_{2} - cos\phi_{1}) = r \cdot sin(\phi_{1}) \cdot
\Delta\phi$. If we assume maser emission could be generated and amplified
through the two maser clouds along the LOS direction and the maser
intensity is proportional to the gain length by multiplying a
constant, g, then the maser flux is F = g $\cdot \Delta$L = g $\cdot r
\cdot sin(\phi_{1}) \cdot \Delta\phi$. Replacing $\Delta\phi$ from
above, finally we have F = g $\cdot r \cdot sin(\phi_{1}) \cdot
\Delta\phi$ = g $\cdot tan(\phi_{1}) \cdot$ r$^{1.5} / \sqrt {GM}$.

\clearpage

\clearpage

\begin{deluxetable}{lcccccc}
\rotate
\tablecolumns{7} \tablewidth{0pc}  
\tablecaption{VLBI Observation Details}
\tablehead {
\colhead{Project Code} & \colhead{Date} &  \colhead{Antennas } &  \colhead{Synthesized Beam} &  \colhead{Track Length} &  \colhead{Sensitivity} &  \colhead{Observing Mode}  
\\
  \colhead{}        &  \colhead{} & \colhead{} & \colhead{(mas x mas, deg)} & \colhead{hour} & \colhead{mJy} & \colhead{}  
            }
\startdata
  BB313 J	   &	2012 Apr 14    	& VLBA+GB+EB  	& 1.04 x 0.27,-10.49  	& 10 		&1.10  		& Self-cal. 		\cr
  BB313 AC  &	2012 Nov 09	& VLBA+GB+EB  	& 1.00 x 0.26,-11.94  	& 12 		&1.25  		& Phase-ref. 		\cr  	  
  BB321 C	   &	2013 Feb 16 	& VLBA+Y	  	&1.42 x 0.47,-16.66  		& 6 		&0.72  		& Self-cal. 		\cr 
  BB321 E1  &	2013 Feb 17 	& VLBA+Y	  	&1.37 x 0.40,-18.05  		& 7 		&0.78  		& Self-cal. 		\cr
  BB321 F	  &	2013 Feb 21 	& VLBA+Y	  	&1.97 x 0.56,-16.08  		& 6.25 	&0.93  		& Self-cal. 		\cr 	
  BB321 L	  &	2013 Mar 13 	& VLBA+GB+Y	  	&-----------------------  		& 7 		&------  		& Self-cal. 		\cr	
  BB321 N	  &	2013 Mar 14 	& VLBA+GB+Y	  	&1.90 x 0.49,-6.68  		& 5 		&0.59  		& Self-cal. 		\cr	
  BB321 S	  &	2013 Apr 13 	& VLBA+GB+Y	  	&-----------------------  		& 7 		&------  		& Self-cal. 		\cr	
  BB321 T	  &	2013 Dec 03 	& VLBA+GB+Y	  	&1.00 x 0.32, -8.93  		& 6 		&0.83  		& Self-cal. 		\cr 	
  BB321 Y2  &	2014 Jan 12 	& VLBA+GB+Y	  	&1.49 x 0.36,-14.00  		& 6 		&0.52  		& Self-cal. 		\cr	
  BB321 Y3  &	2014 Jan 13 	& VLBA+GB+Y	  	&1.52 x 0.38,-12.33  		& 6 		&0.53  		& Self-cal. 		\cr	
  BB321 Y4  &	2014 Jan 16 	& VLBA+GB+Y	  	&1.51 x 0.41,-9.55  		& 6 		&0.42  		& Self-cal. 		\cr	
  BB321 Y5  &	2014 Jan 17 	& VLBA		  	&0.69 x 0.23,-13.19  		& 6 		&2.00  		& Phase-ref. 		\cr %not verified	
  
\enddata
\tablecomments {VLBI observation details. Data for BB321L and BB321S were affected by bad weather and instrument problems, so we didn't make the VLBI map from these 2 tracks.}
\label{table:VLBI observation}
\end{deluxetable}

\begin{deluxetable}{ccccc}
\tablecolumns{5} \tablewidth{0pc}  
\tablecaption{VLBI position for NGC 5765b and the phase-reference calibrators}
\tablehead {
\colhead{Name} & \colhead{Right Ascension} &  \colhead{Declination} &  \colhead{uncertainty in R.A.} &  \colhead{Uncertainty in DEC.}
\\
  \colhead{}        &  \colhead{} & \colhead{} & \colhead{(mas)} & \colhead{(mas)}
            }
            
\startdata
  J1448+0402	&$14^h48^m50^s.361110$	&$+04^{\circ}$02'19".89244	&	1.68	&	2.05	\cr
  J1458+0416	&$14^h58^m59^s.356215$	&$+04^{\circ}$16'13".82056	&	0.03	&	0.05	\cr
  NGC 5765b	&$14^h50^m51^s.51950$	       	&$+05^{\circ}$06'52".2502	&	0.1	&	0.7	\cr

\enddata
\tablecomments {Positions for J1448+0402 and J1458+0416 are from the VLBA Calibrator Survey. Position for NGC 5765b is measured in reference to J1458+0416.}
\label{table:VLBI source position}
\end{deluxetable}

\begin{deluxetable}{ccccccc}
\tablecolumns{6} \tablewidth{0pc}  
\tablecaption{GBT Monitoring Observation Details}
\tablehead {
\colhead{Epoch} & \colhead{Date Observed} &  \colhead{Day Number} &  \colhead{Tsys (K)} &  \colhead{rms noise (mJy)}   
            }
\startdata
  1  &  2012 Feb 26  &    0    &  53.0  &  1.93  		\cr
  2  &  2012 Apr 08  &    42  &  40.0  &  1.23  		\cr
  3  &  2012 May 11  &    75  &  48.3  &  1.38  		\cr
  4  &  2012 Oct 20  &  237  &  48.9  &  1.33  		\cr
  5  &  2012 Nov 13  &  261  &  36.2  &  1.15  		\cr
  6  &  2012 Dec 13  &  291  &  35.7  &  1.07  		\cr
  7  &  2013 Jan 08  &  317  &  38.5  &  1.18  		\cr
  8  &  2013 Feb 25  &  365  &  35.4  &  1.71  		\cr
  9  &  2013 Mar 04  &  372  &  39.1  &  1.21  		\cr
10  &  2013 Apr 06  &  405  &  46.1  &  1.38  		\cr
11  &  2013 May 03  &  432  &  42.4  &  1.17  		\cr
12  &  2013 May 25  &  454  &  41.1  &  1.38  		\cr
13  &  2013 Nov 23  &  636  &  46.4  &  1.65  		\cr
14  &  2013 Dec 24  &  667  &  45.0  &  1.38  		\cr
15$^*$  &  2014 Jan 03  &  677  &  -----   &  1.29  	\cr
16  &  2014 Feb 09  &  724  &  63.5  &  2.18  		\cr
17  &  2013 Mar 08  &  741  &  53.5  &  1.64  		\cr
18  &  2014 Apr 06  &  770  &  38.0  &  1.05  		\cr
19  &  2014 May 09  &  803  &  60.9  &  1.69  		\cr    
\enddata
\tablecomments {$^*$Due to the lack of a GBT observation, here we use the spectrum from the combined VLBI data of BB321T Y2 Y3 \& Y4.}
\label{table:GBT observation}
\end{deluxetable}

\begin{deluxetable}{cccccccc}
\tablecolumns{5} \tablewidth{0pc}  
\tablecaption{Acceleration Measurements for the Systemic Masers from Method 1}
\tablehead {
\colhead{Velocity} & \colhead{Epochs} &  \colhead{Line width} &  \colhead{Acceleration} &  \colhead{A$_\sigma$} 
\\
\colhead{km~s$^{-1}$}  & \colhead{} & \colhead{km~s$^{-1}$} & \colhead{km~s$^{-1}$ yr$^{-1}$} & \colhead{km~s$^{-1}$ yr$^{-1}$} 
            }
\startdata
8261.36  &  1-19  &  2.52  &  0.48  &  0.018		\cr
8264.05  &  1-19  &  2.02  &  0.54  &  0.014		\cr
8267.48  &  1-19  &  3.14  &  0.57  &  0.025		\cr
8269.31  &  1-19  &  2.25  &  0.64  &  0.011		\cr
8271.96  &  1-19  &  3.49  &  0.50  &  0.015		\cr
8275.06  &  1-19  &  2.08  &  0.66  &  0.009		\cr
8277.74  &  1-19  &  2.55  &  0.60  &  0.010		\cr
8280.81  &  1-19  &  2.97  &  0.46  &  0.031		\cr
8283.88  &  1-19  &  3.24  &  0.37  &  0.033		\cr
8286.92  &  1-19  &  2.43  &  0.51  &  0.022		\cr

8291.61  &  4-19  &  3.33  &  0.62  &  0.041		\cr
8293.12  &  4-19  &  1.72  &  0.91  &  0.020		\cr
8294.91  &  4-19  &  2.75  &  0.92  &  0.026		\cr
8297.61  &  4-19  &  3.05  &  0.84  &  0.039		\cr
8299.79  &  4-19  &  2.46  &  0.75  &  0.036		\cr
8302.12  &  4-19  &  2.37  &  0.61  &  0.048		\cr
8303.95  &  4-19  &  2.05  &  0.61  &  0.054		\cr
8305.70  &  4-19  &  2.09  &  0.43  &  0.068		\cr
8307.81  &  4-19  &  2.41  &  0.44  &  0.063		\cr
8309.84  &  4-19  &  2.33  &  0.55  &  0.101		\cr
8311.42  &  4-19  &  2.30  &  0.52  &  0.096		\cr
8313.23  &  4-19  &  2.25  &  0.82  &  0.084		\cr
8315.25  &  4-19  &  2.15  &  0.45  &  0.117		\cr
8317.04  &  4-19  &  2.02  &  0.23  &  0.104		\cr

8318.75  &  4-12  &  1.81  &  1.12  &  0.063		\cr
8320.76  &  4-12  &  1.83  &  0.98  &  0.064		\cr
8323.24  &  4-12  &  2.98  &  0.76  &  0.167		\cr
8326.08  &  4-12  &  1.92  &  1.30  &  0.060		\cr
8328.69  &  4-12  &  3.04  &  0.52  &  0.310		\cr			
\enddata
%\tablecomments {comments}
\label{table:accer-result-method1}
\end{deluxetable}

\begin{deluxetable}{ccc}
\tablecolumns{3} \tablewidth{0pc}  
\tablecaption{Acceleration Measurement for the High Velocity Masers from the ``by eye'' method}
\tablehead {
\colhead{Velocity}  &  \colhead{Accelertion} &  \colhead{A$_\sigma$} 
\\
\colhead{km~s$^{-1}$}  & \colhead{km~s$^{-1}$ yr$^{-1}$} & \colhead{km~s$^{-1}$ yr$^{-1}$} 
            }
\startdata
7539.46	&	 0.100 	&	 0.074		\cr
7549.10	&	-0.175 	&	 0.066		\cr
7582.95	&	-0.528 	&	 0.110		\cr
7690.85	&	 0.222 	&	 0.110		\cr
7702.90	&	 0.006 	&	 0.157		\cr
7723.64	&	 0.007 	&	 0.215		\cr
7729.82	&	 0.269 	&	 0.252		\cr
7748.40	&	 0.088 	&	 0.129		\cr
7769.05	&	 0.458 	&	 0.679		\cr
7772.26	&	-0.038 	&	 0.255		\cr
7795.54	&	 0.282 	&	 0.567		\cr
\tableline
8727.88	&	-0.159	&	0.250		\cr
8730.99	&	 0.199	&	0.095		\cr
8742.55	&	-0.186	&	0.101		\cr
8746.26	&	0.015	&	0.012		\cr
8753.59	&	0.034 	&	 0.018		\cr
8788.41	&	-0.047 	&	0.022		\cr
8845.42	&	-0.007 	&	 0.078		\cr
8847.21	&	-0.215 	&	0.059		\cr
8856.32	&	 0.330 	&	 0.046		\cr
8858.49	&	 0.121 	&	 0.148		\cr
8868.63	&	 0.034 	&	 0.088		\cr
8924.27	&	 0.060 	&	 0.238		\cr
8954.25	&	-0.102 	&	 0.031		\cr
8967.54	&	 0.123 	&	 0.113		\cr
8968.26	&	 0.007 	&	 0.691		\cr
8972.02	&	-0.285 	&	 0.432		\cr

\enddata
%\tablecomments {comments}
\label{table:accer-result-highvelo}
\end{deluxetable}

\begin{deluxetable}{cccccccc}
\tablecolumns{7} \tablewidth{0pc}  
\tablecaption{Input Data for the Disk Fitting Program}
\tablehead {
\colhead{Velocity} & \colhead{X} &  \colhead{X$_\sigma$} &  \colhead{Y} &  \colhead{Y$_\sigma$} &  \colhead{A}  &  \colhead{A$_\sigma$}
\\
\colhead{km~s$^{-1}$}  & \colhead{mas} & \colhead{mas} & \colhead{mas} & \colhead{mas} & \colhead{km~s$^{-1}$ yr$^{-1}$} & \colhead{km~s$^{-1}$ yr$^{-1}$}
            }
\startdata
8345.76   &    0.009   &   0.0143   &   -0.026   &   0.0334   &    1.000   &   1.000   \cr
8344.03   &    0.019   &   0.0071   &   -0.055   &   0.0188   &    1.000   &   1.000   \cr
8342.25   &    0.026   &   0.0056   &   -0.025   &   0.0152   &    1.000   &   1.000   \cr
8340.47   &    0.023   &   0.0048   &   -0.060   &   0.0129   &    1.000   &   1.000   \cr
8338.69   &    0.023   &   0.0042   &   -0.044   &   0.0115   &    1.000   &   1.000   \cr
8336.90   &    0.030   &   0.0060   &   -0.091   &   0.0153   &    1.000   &   1.000   \cr
8335.12   &    0.028   &   0.0047   &   -0.041   &   0.0125   &    1.000   &   1.000   \cr
8333.34   &    0.038   &   0.0029   &   -0.039   &   0.0073   &    1.000   &   1.000   \cr
8331.56   &    0.029   &   0.0032   &   -0.027   &   0.0080   &    1.000   &   1.000   \cr
8329.78   &    0.012   &   0.0036   &   -0.003   &   0.0099   &    0.520   &   0.310   \cr
8327.99   &    0.009   &   0.0019   &   -0.014   &   0.0052   &    0.520   &   0.310   \cr
8326.21   &    0.014   &   0.0020   &   -0.023   &   0.0051   &    1.300   &   0.152   \cr
8324.43   &    0.023   &   0.0029   &   -0.012   &   0.0078   &    0.760   &   0.318   \cr
8322.65   &    0.033   &   0.0020   &   -0.029   &   0.0055   &    0.760   &   0.318   \cr
8320.86   &    0.041   &   0.0012   &   -0.019   &   0.0032   &    0.980   &   0.127   \cr
8319.08   &    0.035   &   0.0016   &   -0.014   &   0.0040   &    1.120   &   0.070   \cr
8317.30   &    0.037   &   0.0017   &   -0.025   &   0.0047   &    0.230   &   0.846   \cr
8315.52   &    0.034   &   0.0012   &   -0.008   &   0.0032   &    0.450   &   0.223   \cr
8313.73   &    0.030   &   0.0011   &   -0.010   &   0.0030   &    0.820   &   0.199   \cr
8311.95   &    0.032   &   0.0011   &   -0.009   &   0.0029   &    0.520   &   0.067   \cr
8310.17   &    0.029   &   0.0010   &   -0.004   &   0.0027   &    0.550   &   0.697   \cr
8308.39   &    0.016   &   0.0011   &    0.003   &   0.0029   &    0.440   &   0.287   \cr
8306.61   &    0.000   &   0.0010   &    0.011   &   0.0028   &    0.430   &   0.105   \cr
8304.82   &    0.000   &   0.0009   &    0.000   &   0.0025   &    0.610   &   0.186   \cr
8303.04   &   -0.002   &   0.0011   &    0.012   &   0.0028   &    0.610   &   0.186   \cr
8301.26   &   -0.002   &   0.0008   &    0.010   &   0.0021   &    0.750   &   0.243   \cr
8299.48   &   -0.003   &   0.0007   &    0.009   &   0.0019   &    0.750   &   0.243   \cr
8297.69   &   -0.002   &   0.0007   &    0.010   &   0.0019   &    0.840   &   0.233   \cr
8295.91   &   -0.002   &   0.0007   &    0.013   &   0.0020   &    0.920   &   0.075   \cr
8294.13   &   -0.012   &   0.0008   &    0.018   &   0.0021   &    0.920   &   0.075   \cr
\tablebreak
8292.35   &   -0.016   &   0.0010   &    0.016   &   0.0027   &    0.620   &   0.065   \cr
8290.56   &   -0.024   &   0.0015   &    0.021   &   0.0039   &    0.620   &   0.065   \cr
8288.78   &   -0.031   &   0.0016   &    0.023   &   0.0042   &    0.510   &   0.046   \cr
8287.00   &   -0.037   &   0.0013   &    0.030   &   0.0034   &    0.510   &   0.046   \cr
8285.22   &   -0.040   &   0.0014   &    0.029   &   0.0036   &    0.370   &   0.045   \cr
8283.44   &   -0.044   &   0.0010   &    0.029   &   0.0028   &    0.370   &   0.045   \cr
8281.65   &   -0.049   &   0.0010   &    0.043   &   0.0027   &    0.460   &   0.033   \cr
8279.87   &   -0.053   &   0.0009   &    0.042   &   0.0024   &    0.600   &   0.010   \cr
8278.09   &   -0.052   &   0.0008   &    0.041   &   0.0022   &    0.600   &   0.010   \cr
8276.31   &   -0.056   &   0.0009   &    0.043   &   0.0024   &    0.660   &   0.022   \cr
8274.53   &   -0.060   &   0.0009   &    0.048   &   0.0024   &    0.500   &   0.062   \cr
8272.74   &   -0.062   &   0.0008   &    0.049   &   0.0022   &    0.500   &   0.062   \cr
8270.96   &   -0.063   &   0.0008   &    0.048   &   0.0022   &    0.500   &   0.062   \cr
8269.18   &   -0.065   &   0.0008   &    0.049   &   0.0021   &    0.640   &   0.071   \cr
8267.40   &   -0.064   &   0.0010   &    0.056   &   0.0025   &    0.570   &   0.123   \cr
8265.61   &   -0.065   &   0.0012   &    0.057   &   0.0033   &    0.540   &   0.052   \cr
8263.83   &   -0.064   &   0.0013   &    0.065   &   0.0034   &    0.540   &   0.052   \cr
8262.05   &   -0.063   &   0.0013   &    0.073   &   0.0034   &    0.480   &   0.053   \cr
8260.27   &   -0.058   &   0.0020   &    0.058   &   0.0051   &    0.480   &   0.053   \cr
8258.48   &   -0.078   &   0.0028   &    0.084   &   0.0074   &    1.000   &   1.000   \cr
8256.70   &   -0.075   &   0.0028   &    0.069   &   0.0075   &    1.000   &   1.000   \cr
8254.92   &   -0.083   &   0.0025   &    0.082   &   0.0065   &    1.000   &   1.000   \cr
8253.14   &   -0.088   &   0.0024   &    0.091   &   0.0065   &    1.000   &   1.000   \cr
8251.36   &   -0.089   &   0.0020   &    0.102   &   0.0053   &    1.000   &   1.000   \cr
8249.57   &   -0.086   &   0.0021   &    0.091   &   0.0056   &    1.000   &   1.000   \cr
8247.79   &   -0.090   &   0.0019   &    0.092   &   0.0053   &    1.000   &   1.000   \cr
8246.01   &   -0.090   &   0.0027   &    0.109   &   0.0074   &    1.000   &   1.000   \cr
8244.23   &   -0.093   &   0.0042   &    0.112   &   0.0116   &    1.000   &   1.000   \cr
8242.44   &   -0.077   &   0.0051   &    0.089   &   0.0140   &    1.000   &   1.000   \cr
8240.66   &   -0.095   &   0.0075   &    0.111   &   0.0204   &    1.000   &   1.000   \cr
\tablebreak
9081.29   &    0.297   &   0.0163   &   -0.611   &   0.0474   &    0.000   &   0.200   \cr
9079.50   &    0.273   &   0.0150   &   -0.701   &   0.0457   &    0.000   &   0.200   \cr
9077.71   &    0.274   &   0.0176   &   -0.642   &   0.0533   &    0.000   &   0.200   \cr
8973.79   &    0.456   &   0.0253   &   -0.960   &   0.0589   &    0.000   &   0.200   \cr
8972.08   &    0.436   &   0.0100   &   -0.895   &   0.0300   &   -0.285   &   0.476   \cr
8970.29   &    0.419   &   0.0100   &   -0.798   &   0.0300   &    0.000   &   0.200   \cr
8968.50   &    0.431   &   0.0100   &   -0.820   &   0.0300   &    0.007   &   0.719   \cr
8966.71   &    0.422   &   0.0100   &   -0.837   &   0.0300   &    0.123   &   0.230   \cr
8964.92   &    0.429   &   0.0100   &   -0.835   &   0.0300   &    0.000   &   0.200   \cr
8963.13   &    0.437   &   0.0104   &   -0.853   &   0.0300   &    0.000   &   0.200   \cr
8961.35   &    0.417   &   0.0100   &   -0.851   &   0.0300   &    0.000   &   0.200   \cr
8959.56   &    0.458   &   0.0100   &   -0.897   &   0.0300   &    0.000   &   0.200   \cr
8957.77   &    0.432   &   0.0100   &   -0.817   &   0.0300   &    0.000   &   0.200   \cr
8955.98   &    0.421   &   0.0100   &   -0.841   &   0.0300   &    0.000   &   0.200   \cr
8954.19   &    0.435   &   0.0100   &   -0.871   &   0.0300   &   -0.102   &   0.202   \cr
8952.40   &    0.409   &   0.0100   &   -0.833   &   0.0300   &    0.000   &   0.200   \cr
8950.61   &    0.432   &   0.0137   &   -0.854   &   0.0353   &    0.000   &   0.200   \cr
8948.75   &    0.384   &   0.0272   &   -0.752   &   0.0637   &    0.000   &   0.200   \cr
8946.97   &    0.423   &   0.0266   &   -0.744   &   0.0622   &    0.000   &   0.200   \cr
8927.43   &    0.435   &   0.0164   &   -0.838   &   0.0541   &    0.000   &   0.200   \cr
8925.57   &    0.461   &   0.0114   &   -0.835   &   0.0300   &    0.000   &   0.200   \cr
8923.78   &    0.478   &   0.0100   &   -0.851   &   0.0300   &    0.060   &   0.311   \cr
8921.99   &    0.488   &   0.0128   &   -0.880   &   0.0378   &    0.000   &   0.200   \cr
8920.21   &    0.483   &   0.0138   &   -0.924   &   0.0373   &    0.000   &   0.200   \cr
8918.42   &    0.498   &   0.0100   &   -0.945   &   0.0300   &    0.000   &   0.200   \cr
8916.63   &    0.488   &   0.0100   &   -0.924   &   0.0300   &    0.000   &   0.200   \cr
8914.84   &    0.480   &   0.0100   &   -0.933   &   0.0300   &    0.000   &   0.200   \cr
8913.05   &    0.502   &   0.0100   &   -0.933   &   0.0300   &    0.000   &   0.200   \cr
8911.26   &    0.504   &   0.0100   &   -0.914   &   0.0300   &    0.000   &   0.200   \cr
8909.47   &    0.514   &   0.0100   &   -0.996   &   0.0300   &    0.000   &   0.200   \cr
\tablebreak
8904.11   &    0.502   &   0.0139   &   -0.912   &   0.0340   &    0.000   &   0.200   \cr
8888.07   &    0.481   &   0.0143   &   -1.056   &   0.0428   &    0.000   &   0.200   \cr
8880.79   &    0.497   &   0.0270   &   -0.862   &   0.0588   &    0.000   &   0.200   \cr
8875.43   &    0.572   &   0.0281   &   -1.111   &   0.0657   &    0.000   &   0.200   \cr
8873.70   &    0.594   &   0.0131   &   -0.991   &   0.0337   &    0.000   &   0.200   \cr
8871.91   &    0.603   &   0.0118   &   -1.036   &   0.0306   &    0.000   &   0.200   \cr
8870.12   &    0.612   &   0.0106   &   -1.048   &   0.0300   &    0.000   &   0.200   \cr
8868.33   &    0.615   &   0.0100   &   -1.055   &   0.0300   &    0.034   &   0.219   \cr
8866.54   &    0.609   &   0.0100   &   -1.074   &   0.0300   &    0.000   &   0.200   \cr
8864.75   &    0.653   &   0.0124   &   -1.052   &   0.0331   &    0.000   &   0.200   \cr
8862.96   &    0.604   &   0.0110   &   -1.024   &   0.0300   &    0.000   &   0.200   \cr
8861.17   &    0.646   &   0.0100   &   -1.116   &   0.0300   &    0.000   &   0.200   \cr
8859.39   &    0.648   &   0.0100   &   -1.100   &   0.0300   &    0.121   &   0.249   \cr
8857.60   &    0.654   &   0.0100   &   -1.110   &   0.0300   &    0.121   &   0.249   \cr
8855.81   &    0.658   &   0.0100   &   -1.108   &   0.0300   &    0.330   &   0.205   \cr
8854.02   &    0.662   &   0.0100   &   -1.123   &   0.0300   &    0.000   &   0.200   \cr
8852.23   &    0.652   &   0.0100   &   -1.108   &   0.0300   &    0.000   &   0.200   \cr
8850.44   &    0.660   &   0.0100   &   -1.111   &   0.0300   &    0.000   &   0.200   \cr
8848.65   &    0.670   &   0.0100   &   -1.118   &   0.0300   &    0.000   &   0.200   \cr
8846.87   &    0.672   &   0.0100   &   -1.136   &   0.0300   &   -0.215   &   0.209   \cr
8845.08   &    0.675   &   0.0100   &   -1.135   &   0.0300   &   -0.007   &   0.215   \cr
8843.29   &    0.679   &   0.0100   &   -1.121   &   0.0300   &    0.000   &   0.200   \cr
8841.50   &    0.711   &   0.0100   &   -1.178   &   0.0300   &    0.000   &   0.200   \cr
8839.71   &    0.686   &   0.0115   &   -1.160   &   0.0300   &    0.000   &   0.200   \cr
8837.87   &    0.634   &   0.0210   &   -1.152   &   0.0492   &    0.000   &   0.200   \cr
8789.62   &    0.817   &   0.0118   &   -1.263   &   0.0324   &    0.000   &   0.200   \cr
8787.84   &    0.855   &   0.0100   &   -1.342   &   0.0300   &   -0.047   &   0.201   \cr
8786.05   &    0.862   &   0.0100   &   -1.383   &   0.0300   &    0.000   &   0.200   \cr
8784.26   &    0.863   &   0.0100   &   -1.368   &   0.0300   &    0.000   &   0.200   \cr
8782.43   &    0.893   &   0.0231   &   -1.436   &   0.0541   &    0.000   &   0.200   \cr
\tablebreak
8759.21   &    1.030   &   0.0140   &   -1.527   &   0.0393   &    0.000   &   0.200   \cr
8757.43   &    1.038   &   0.0100   &   -1.514   &   0.0300   &    0.000   &   0.200   \cr
8755.64   &    1.043   &   0.0100   &   -1.521   &   0.0300   &    0.000   &   0.200   \cr
8753.85   &    1.040   &   0.0100   &   -1.509   &   0.0300   &    0.034   &   0.201   \cr
8752.06   &    1.044   &   0.0100   &   -1.516   &   0.0300   &    0.000   &   0.200   \cr
8750.27   &    1.054   &   0.0100   &   -1.538   &   0.0300   &    0.000   &   0.200   \cr
8748.48   &    1.049   &   0.0126   &   -1.499   &   0.0345   &    0.000   &   0.200   \cr
8746.69   &    1.078   &   0.0100   &   -1.543   &   0.0300   &    0.015   &   0.200   \cr
8744.90   &    1.086   &   0.0100   &   -1.558   &   0.0300   &    0.000   &   0.200   \cr
8743.12   &    1.092   &   0.0100   &   -1.537   &   0.0300   &   -0.186   &   0.224   \cr
8741.33   &    1.087   &   0.0100   &   -1.549   &   0.0300   &    0.000   &   0.200   \cr
8730.62   &    1.161   &   0.0182   &   -1.561   &   0.0627   &    0.199   &   0.221   \cr
8727.04   &    1.190   &   0.0177   &   -1.474   &   0.0499   &   -0.159   &   0.320   \cr
7796.67   &   -0.697   &   0.0100   &    0.928   &   0.0300   &    0.282   &   0.601   \cr
7794.89   &   -0.698   &   0.0100   &    0.901   &   0.0300   &    0.282   &   0.601   \cr
7793.12   &   -0.680   &   0.0160   &    0.818   &   0.0460   &    0.000   &   0.200   \cr
7789.57   &   -0.664   &   0.0151   &    0.806   &   0.0409   &    0.000   &   0.200   \cr
7787.79   &   -0.665   &   0.0154   &    0.794   &   0.0368   &    0.000   &   0.200   \cr
7786.02   &   -0.697   &   0.0108   &    0.857   &   0.0325   &    0.000   &   0.200   \cr
7784.24   &   -0.678   &   0.0100   &    0.821   &   0.0300   &    0.000   &   0.200   \cr
7782.47   &   -0.699   &   0.0100   &    0.868   &   0.0300   &    0.000   &   0.200   \cr
7780.69   &   -0.670   &   0.0132   &    0.800   &   0.0354   &    0.000   &   0.200   \cr
7777.21   &   -0.666   &   0.0174   &    0.757   &   0.0505   &    0.000   &   0.200   \cr
7775.43   &   -0.642   &   0.0147   &    0.816   &   0.0433   &    0.000   &   0.200   \cr
7773.59   &   -0.657   &   0.0100   &    0.758   &   0.0300   &    0.000   &   0.200   \cr
7771.82   &   -0.664   &   0.0100   &    0.804   &   0.0300   &   -0.038   &   0.324   \cr
7770.04   &   -0.652   &   0.0100   &    0.788   &   0.0300   &    0.000   &   0.200   \cr
7768.27   &   -0.656   &   0.0100   &    0.795   &   0.0300   &    0.458   &   0.708   \cr
7766.49   &   -0.639   &   0.0100   &    0.794   &   0.0300   &    0.000   &   0.200   \cr
7764.72   &   -0.602   &   0.0106   &    0.759   &   0.0300   &    0.000   &   0.200   \cr
\tablebreak
7762.88   &   -0.646   &   0.0254   &    0.762   &   0.0654   &    0.000   &   0.200   \cr
7761.17   &   -0.605   &   0.0127   &    0.745   &   0.0324   &    0.000   &   0.200   \cr
7759.45   &   -0.627   &   0.0172   &    0.687   &   0.0483   &    0.000   &   0.200   \cr
7757.62   &   -0.595   &   0.0100   &    0.690   &   0.0300   &    0.000   &   0.200   \cr
7755.84   &   -0.625   &   0.0124   &    0.756   &   0.0328   &    0.000   &   0.200   \cr
7754.07   &   -0.612   &   0.0100   &    0.794   &   0.0300   &    0.000   &   0.200   \cr
7752.29   &   -0.592   &   0.0100   &    0.749   &   0.0300   &    0.000   &   0.200   \cr
7750.52   &   -0.601   &   0.0100   &    0.751   &   0.0300   &    0.000   &   0.200   \cr
7748.74   &   -0.576   &   0.0100   &    0.719   &   0.0300   &    0.088   &   0.238   \cr
7746.97   &   -0.582   &   0.0100   &    0.710   &   0.0300   &    0.000   &   0.200   \cr
7745.14   &   -0.552   &   0.0230   &    0.712   &   0.0538   &    0.000   &   0.200   \cr
7743.42   &   -0.563   &   0.0100   &    0.740   &   0.0300   &    0.000   &   0.200   \cr
7741.64   &   -0.538   &   0.0100   &    0.746   &   0.0300   &    0.000   &   0.200   \cr
7739.87   &   -0.541   &   0.0104   &    0.701   &   0.0300   &    0.000   &   0.200   \cr
7738.09   &   -0.547   &   0.0113   &    0.702   &   0.0313   &    0.000   &   0.200   \cr
7736.32   &   -0.533   &   0.0116   &    0.674   &   0.0316   &    0.000   &   0.200   \cr
7734.54   &   -0.495   &   0.0131   &    0.583   &   0.0341   &    0.000   &   0.200   \cr
7732.77   &   -0.497   &   0.0100   &    0.662   &   0.0300   &    0.000   &   0.200   \cr
7730.99   &   -0.529   &   0.0100   &    0.672   &   0.0300   &    0.000   &   0.200   \cr
7729.22   &   -0.534   &   0.0100   &    0.690   &   0.0300   &    0.269   &   0.322   \cr
7727.44   &   -0.526   &   0.0100   &    0.666   &   0.0300   &    0.000   &   0.200   \cr
7725.67   &   -0.509   &   0.0100   &    0.650   &   0.0300   &    0.000   &   0.200   \cr
7723.89   &   -0.534   &   0.0100   &    0.711   &   0.0300   &    0.007   &   0.294   \cr
7722.12   &   -0.508   &   0.0100   &    0.619   &   0.0300   &    0.000   &   0.200   \cr
7720.29   &   -0.480   &   0.0255   &    0.611   &   0.0596   &    0.000   &   0.200   \cr
7702.54   &   -0.413   &   0.0236   &    0.509   &   0.0597   &    0.006   &   0.254   \cr
7693.76   &   -0.454   &   0.0100   &    0.582   &   0.0300   &    0.000   &   0.200   \cr
7691.94   &   -0.453   &   0.0100   &    0.580   &   0.0300   &    0.000   &   0.200   \cr
7690.17   &   -0.457   &   0.0100   &    0.598   &   0.0300   &    0.222   &   0.228   \cr
7688.39   &   -0.454   &   0.0100   &    0.530   &   0.0300   &    0.000   &   0.200   \cr
\tablebreak
7686.66   &   -0.469   &   0.0169   &    0.575   &   0.0472   &    0.000   &   0.200   \cr
7684.84   &   -0.445   &   0.0122   &    0.599   &   0.0336   &    0.000   &   0.200   \cr
7683.11   &   -0.437   &   0.0117   &    0.604   &   0.0339   &    0.000   &   0.200   \cr
7681.29   &   -0.444   &   0.0128   &    0.591   &   0.0351   &    0.000   &   0.200   \cr
7626.26   &   -0.368   &   0.0100   &    0.448   &   0.0300   &    0.000   &   0.200   \cr
7624.49   &   -0.362   &   0.0100   &    0.473   &   0.0300   &    0.000   &   0.200   \cr
7622.71   &   -0.354   &   0.0114   &    0.450   &   0.0322   &    0.000   &   0.200   \cr
7620.94   &   -0.364   &   0.0134   &    0.528   &   0.0343   &    0.000   &   0.200   \cr
7619.16   &   -0.360   &   0.0148   &    0.461   &   0.0378   &    0.000   &   0.200   \cr
7601.44   &   -0.376   &   0.0140   &    0.508   &   0.0406   &    0.000   &   0.200   \cr
7599.67   &   -0.344   &   0.0165   &    0.408   &   0.0480   &    0.000   &   0.200   \cr
7597.83   &   -0.379   &   0.0214   &    0.416   &   0.0500   &    0.000   &   0.200   \cr
7596.06   &   -0.400   &   0.0272   &    0.418   &   0.0636   &    0.000   &   0.200   \cr
7592.54   &   -0.380   &   0.0132   &    0.544   &   0.0361   &    0.000   &   0.200   \cr
7590.76   &   -0.332   &   0.0115   &    0.344   &   0.0306   &    0.000   &   0.200   \cr
7588.99   &   -0.347   &   0.0100   &    0.396   &   0.0300   &    0.000   &   0.200   \cr
7587.21   &   -0.351   &   0.0105   &    0.421   &   0.0300   &    0.000   &   0.200   \cr
7583.64   &   -0.356   &   0.0100   &    0.411   &   0.0300   &   -0.528   &   0.228   \cr
7581.89   &   -0.356   &   0.0100   &    0.444   &   0.0300   &    0.000   &   0.200   \cr
7578.34   &   -0.356   &   0.0104   &    0.412   &   0.0300   &    0.000   &   0.200   \cr
7576.56   &   -0.344   &   0.0106   &    0.396   &   0.0300   &    0.000   &   0.200   \cr
7574.81   &   -0.342   &   0.0153   &    0.410   &   0.0445   &    0.000   &   0.200   \cr
7572.99   &   -0.388   &   0.0252   &    0.452   &   0.0589   &    0.000   &   0.200   \cr
7555.24   &   -0.336   &   0.0237   &    0.364   &   0.0556   &    0.000   &   0.200   \cr
7549.94   &   -0.336   &   0.0103   &    0.323   &   0.0300   &   -0.175   &   0.211   \cr
7548.16   &   -0.334   &   0.0100   &    0.389   &   0.0300   &    0.000   &   0.200   \cr
7546.41   &   -0.325   &   0.0171   &    0.367   &   0.0497   &    0.000   &   0.200   \cr
7544.63   &   -0.363   &   0.0165   &    0.370   &   0.0478   &    0.000   &   0.200   \cr
7541.08   &   -0.334   &   0.0110   &    0.331   &   0.0319   &    0.000   &   0.200   \cr
7539.30   &   -0.342   &   0.0100   &    0.355   &   0.0300   &    0.100   &   0.213   \cr
7537.53   &   -0.306   &   0.0141   &    0.312   &   0.0409   &    0.000   &   0.200   \cr
7542.85   &   -0.306   &   0.0157   &    0.334   &   0.0455   &    0.000   &   0.200   \cr
\enddata
\tablecomments {Input data for the Bayesian fitting program for NGC 5765b. No error floor for the position data has been added. An error floor of 0.2 km~s$^{-1}$ yr$^{-1}$ was added for the acceleration data of the high-velocity masers.}
\label{table:disk fitting data}
\end{deluxetable}

\begin{deluxetable}{cccccccc}
\tablecolumns{4} \tablewidth{0pc}  
\tablecaption{Disk Fitting Results from NGC5765b}
\tablehead {
\colhead{Parameter} & \colhead{Priors} &  \colhead{Posterioris} &  \colhead{Units}   
            }
\startdata
$H_{0}$	&	...			&	66.0 $\pm$ 5.0			&	km~s$^{-1}$ Mpc$^{-1}$	\cr
M		&	...			&	4.55 $\pm$ 0.31		&	10$^{7}$ M$_{\odot}$		\cr 
V		&	8304 $\pm$ 30	&	8334.60 $\pm$ 1.13		&	km~s$^{-1}$			\cr
$X_{0}$	&	...			&	-0.044 $\pm$ 0.0025		&	mas					\cr
$Y_{0}$	&	...			&	-0.100 $\pm$ 0.0025		&	mas					\cr
$i_{0}$	&	...			&	94.5 $\pm$ 0.25		&	deg					\cr
$di/dr$	&	...			&	-10.6 $\pm$ 1.13		&	deg mas$^{-1}$					\cr
$p_{0}$	&	...			&	146.7 $\pm$ 0.125		&	deg					\cr
$dp/dr$	&	...			&	-3.46 $\pm$ 0.43		&	deg mas$^{-1}$					\cr
$V_{cor}$	& 0 $\pm$ 250		&	2.0 $\pm$ 282			&	km~s$^{-1}$					\cr

\enddata
\tablecomments {Fitting results for the global parameters describing the maser disk, which are Hubble constant ($H_{0}$); black hole mass (M); recession velocity of NGC 5765b, with optical definition (V); position of the dynamic center ($X_{0}$, $Y_{0}$); inclination of the disk ($i_{0}$); warping of the disk on the inclination angle direction ($di/dr$); position angle of the disk ($p_{0}$); warping of the disk on the position angle direction ($dp/dr$); velocity correction due to any peculiar motion of NGC 5765b with respect to the Hubble flow ($V_{cor}$). For each parameter, we quote the mean value in its probability distribution function as the fitted result, and 68\% confidence range as the error. All uncertainties listed here have been scaled by $\sqrt{\chi^{2}/N}$, i.e., $\sqrt{1.575}$. }
\label{table:Bayesian fitting result}
\end{deluxetable}

\begin{deluxetable}{lllll}
\rotate
\tablecolumns{5} \tablewidth{0pc}  
\tablecaption{Table of $\Ho$ uncertainty terms}
\tablehead {
\colhead{Formal uncertainty} & \colhead{} &  \colhead{$\Ho$} &  \colhead{1 $\sigma$ Uncertainty} &  \colhead{}
\\
  \colhead{}        &  \colhead{} & \colhead{(km~s$^{-1}$ Mpc$^{-1}$)} & \colhead{(km~s$^{-1}$ Mpc$^{-1}$)} & \colhead{(\%)} 
            }
\startdata
Base model	   			&	    					& 	66.0 						& $\pm$ 5.0  				&    7.6 		\cr
\tableline
Systematic uncertainties	   	&	Value in Base Model		& 	$\Ho$  					& Difference from base model  	& 			\cr
						&						&	(km~s$^{-1}$ Mpc$^{-1}$)		& (km~s$^{-1}$ Mpc$^{-1}$)	&   (\%)		\cr
\tableline
Different systemic feature\\ 
acceleration error floor		&	...					&	68.0						&	2.0					&	3.00		\cr

Different hv-feature\\ 
acceleration error floor		&	0.2 km~s$^{-1}$ yr$^{-1}$	&	66.5						&	0.5					&	0.75		\cr

Different method for measurement \\
of systemic accelerations  		&...						&68.5					&	2.5					&	3.8		\cr

Unmodeled spiral structure		&...						&...						&    0.66					&	1.0		\cr
\tableline
Systematic uncertainties\\ 
added in quadrature			&						&							&	3.31					&	5.0		\cr
\tableline
Uncertainty in the $\Ho$ based\\ 
on NGC 5765b maser distance	&						&							&	6.0					&	9.1		\cr

\enddata
\tablecomments {The final uncertainty is calculated by adding individual sources of uncertainty in quadrature.}
\label{table:H0uncertainty}
\end{deluxetable}

\begin{deluxetable}{cccc}
\tablecolumns{4} \tablewidth{0pc}  
\tablecaption{Comparison of input data between NGC 4258 and NGC 5765b}
\tablehead {
\colhead{Name} & \colhead{Red-shifted} &  \colhead{Systemic} &  \colhead{Blue-shifted} 
            }
\startdata
  Number of data points						\cr
  NGC 4258	&	151	&	187	&	32	\cr
  NGC 5765b	&	71	&	71	&	70	\cr
 \tableline
  Velocity Range (km~s$^{-1}$)						\cr
  NGC 4258	&	420	&	195 (110)	&	236	\cr
  NGC 5765b	&	354	&	90 (70)		&	253	\cr
\enddata
\tablecomments {Comparison of the input data for the Bayesian fitting between NGC 4258 and NGC 5765b. For the velocity range of systemic masers, the number listed in the brackets are velocity ranges with valid accelerations measured.}
\label{table:datacomparison}
\end{deluxetable}

\end{document}